\documentclass[9pt,conference]{IEEEtran}
\IEEEoverridecommandlockouts
\usepackage{fancyhdr}
\usepackage[normalem]{ulem}
\usepackage{url}
\usepackage{tikz}
\usepackage{mathptmx} % USE Times Fonts
\usepackage{amsthm}
\usepackage{amsmath,amsfonts}
\usepackage{graphicx}
\usepackage{textcomp}
\usepackage[table,xcdraw]{}
\usepackage{booktabs} % For formal tables
\usepackage{stfloats}
\usepackage{subfigure}
\usepackage{multirow}
\usepackage{paralist}
\usepackage{balance}
\usepackage{bm}
\usepackage{pifont}
\usepackage{subcaption}
\usepackage{url}
\usepackage[hidelinks]{hyperref}

\usepackage[bottom=2.1cm, top=1.20cm, left=1.72cm, right=1.72cm]{geometry}
\usepackage{xspace}
\usepackage{algorithmic}
\usepackage[ruled,linesnumbered]{algorithm2e}
\usepackage{diagbox}
\usepackage{threeparttable}
\usepackage{makecell}
\usepackage{colortbl}
\usepackage{bm}
\definecolor{grey}{RGB}{0.5,0.5,0.5}
\usepackage{tcolorbox}
\newcommand{\parlabel}[1]{{\noindent\bf #1}}
\newcommand{\eg}{e.g.\xspace}

\newrobustcmd*\circled[1]{\tikz[baseline=(char.base)]{
            \node[shape=circle,draw,inner sep=1pt,fill,text=white,minimum size=1em] (char) {\textsf{\small #1}};}}

\usepackage{amsmath,amssymb,amsfonts}

\usepackage{cite}
\usepackage{caption}
\usepackage{colortbl}
\usepackage{tabularx}
\usepackage{textcomp}
\definecolor{darkgreen}{rgb}{0, 0.6, 0}

\begin{document}

% {RGB}{66, 186, 71}

\newcommand{\approach}{UVMatic}
\newcommand\name{UVM{$^\text{2}$}}
\renewcommand{\arraystretch}{1.48}
% \begin{document}

% \title{UVMatic: Leveraging LLMs and Scripting for Hierarchical UVM Testbench Generation
% }
\title{From Concept to Practice: an Automated LLM-aided UVM Machine for RTL Verification}
\author{
    \IEEEauthorblockN{
        Junhao Ye$^{1,2}$, Yuchen Hu$^{1,2}$, Ke Xu$^{1,2}$, Dingrong Pan$^2$, Qichun Chen$^{3}$, Jie Zhou$^{1,2}$\\ Shuai Zhao$^{4}$, Xinwei Fang$^{5,*}$, Xi Wang$^{1,2,*}$, Nan Guan$^{6}$, Zhe Jiang$^{1,2,*}$\thanks{*Corresponding authors: xinwei.fang@york.ac.uk, xi.wang@seu.edu.cn, and zhejiang.uk@gmail.com.
       \newline {This paper was accepted by \textbf{ICCAD 2025}.}  {Please cite the published version with DOI:  \textbf{10.1109/ICCAD66269.2025.11240679}.}}
    }

    \IEEEauthorblockA{
        $^1$School of Integrated Circuits, Southeast University, China $^2$National Center of Technology Innovation for EDA, China\\
        $^3$College of Economics, Shenzhen University, China\\
        $^4$School of Computer Science and Engineering, Sun Yat-sen University, China\\
        $^5$Department of Computer Science, University of York, UK\\
        $^6$Department of Computer Science, City University of Hong Kong, Hong Kong\\
    }
    \vspace{-20pt}
}

\maketitle
\begin{abstract}
\label{sc:abs}
% Verification is a critical bottleneck in the development of integrated circuits (IC)\guan{redundant wording: Verification is a critical ... of  integrated circuits development}, consuming nearly 70\% of the total development cycle. 
% This is mainly caused by the complexity of constructing UVM \guan{what is UVM? full name? also the complexity of UVM cannot support the first sentence, as UVM is only one way of verification; the logic should be: verification is bottleneck ---- UVM is a good way for verification because  ... ---- it is still a bottleneck even if using UVM because of the complexity of ... } verification environment and generating adequate stimuli for sufficient tests, which requires extensive coding efforts, in-depth domain expertise to understand the complex design, and repeated engagement with multiple EDA tools through infinity test iterations \guan{infinity? what does it mean?}.
% Here\guan{in this paper}, we present \name, a verification automation framework that leverages large language models (LLMs) to automatically generate UVM-compliant testbench and iteratively refine it based on coverage metrics, minimising manual effort while maintaining comprehensive converge. 
% To validate \name, we introduce a new benchmark suite featuring RTL designs of up to 1.5K lines of code. Results show that \name\ reduces setup time by up to 100× compared to an expert verification engineer and achieves an average coverage of 92\%, outperforming the state-of-the-art by 9×.
Verification presents a major bottleneck in Integrated Circuit (IC) development, consuming nearly 70\% of the total development effort. While the Universal Verification Methodology (UVM) is widely used in industry to improve verification efficiency through structured and reusable testbenches, constructing these testbenches and generating sufficient stimuli remain challenging. 
These challenges arise from the considerable manual coding effort required, repetitive manual execution of multiple EDA tools, and the need for in-depth domain expertise to navigate complex designs.
Here, we present \name, an automated verification framework that leverages Large Language Models (LLMs) to generate UVM testbenches and iteratively refine them using coverage feedback, significantly reducing manual effort while maintaining rigorous verification standards.
To evaluate \name, we introduce a benchmark suite comprising Register Transfer Level (RTL) designs of up to 1.6K lines of code.
The results show that \name\ reduces testbench setup time by up to 38.82× compared to experienced engineers, and achieve average code and function coverage of 87.44\% and 89.58\%, outperforming state-of-the-art solutions by 20.96\% and 23.51\%, respectively.

\end{abstract}

\section{Introduction}
\label{sc:int}

\begin{figure}[t]
    \includegraphics[trim=00mm 00mm 00mm 00mm, clip, width=1\linewidth]{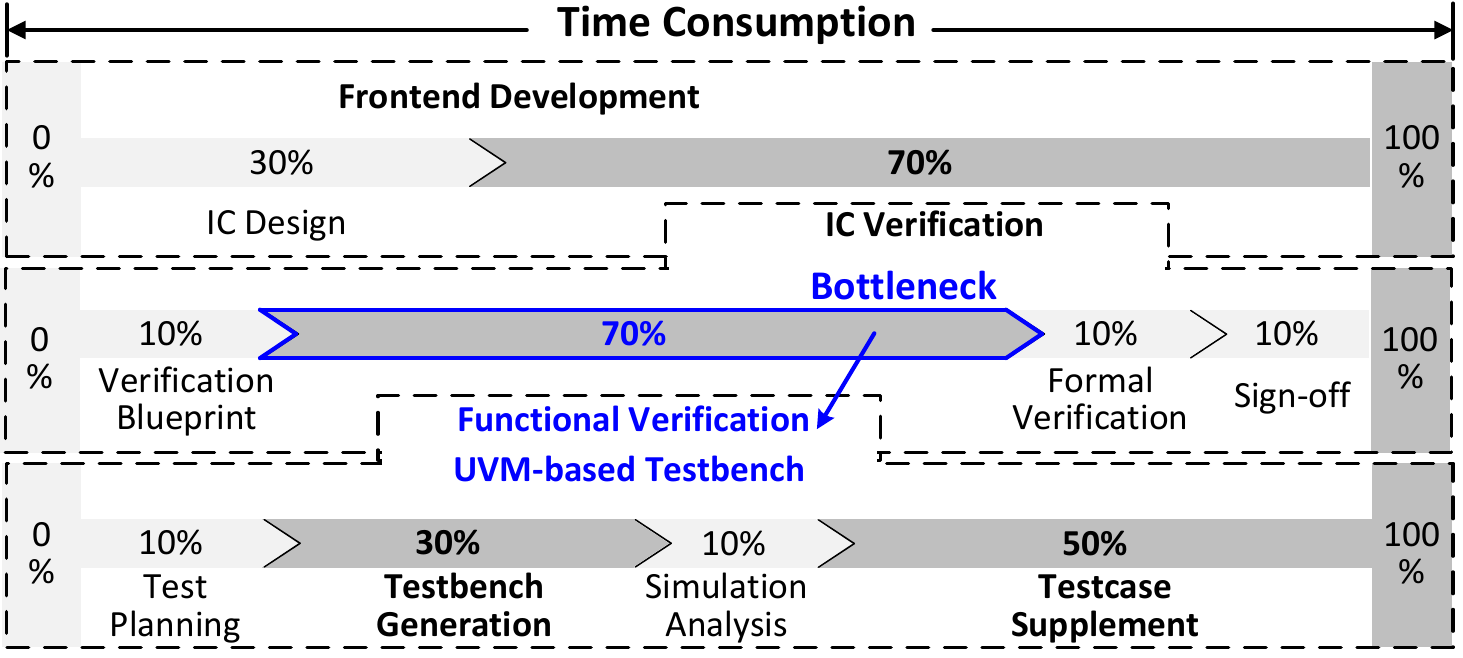} % 可根据需求调整宽度，如 0.9\textwidth

    \caption{
    Breakdown of the IC frontend design and verification workflow, demonstrating the effort required at each stage, where the functional verification dominates the time consumption of the entire workflow. 
    % The frontend design in IC development consists of design and verification, with verification consuming around 70\% of the cycles. For frontend verification, function verification dominates with 70\% of the time, primarily allocated to testbench generation (30\%) and test case supplementation (50\%), underscoring their critical roles in the workflow.
    % \zhenote{Please make the caption longer to describe the meaning of the figure.}
    }
    \vspace{-10pt}
    \label{fig:Intro} % 便于后续引用
\end{figure}

IC verification 
% \guan{veirification or design verification? be consistent}
is the most resource and time-intensive phase in IC frontend design, consuming nearly 70\% of the total development effort~\cite{lahti2018we}. As illustrated in Fig.~\ref{fig:Intro}, 
% \guan{Fig.1 or Fig.? be consistent with the figure; don't abuse bold font}
IC verification usually begins with a verification blueprint, followed by functional verification, formal verification, and ultimately sign-off~\cite{ashar2019closing}. Among these stages, the functional verification alone accounts for approximately 70\% of the total verification effort and encounters two significant challenges that hinder verification efficiency~\cite{ismail2021high,danciu2022coverage,gal2022machine,hennessy2011computer,harris2021digital}.

The first challenge is the complexity associated with testbench generation. UVM-based testbench is widely used in industrial due to its modular, layered architectures and reusable verification components~\cite{salah2014uvm,madan2015pragmatic,vineeth2018uvm}. For instance, stimulus generation is modularly encapsulated via sequencers, enabling reuse across designs~\cite{yun2011beyond,piccolboni2014simplified,visalli2017uvm,vintila2018portable}. However, this modularity significantly increases the complexity of testbench setup. Engineers must manually instantiate and configure numerous hierarchical components adhering to SystemVerilog class conventions and UVM library specifications~\cite{bromley2013if,francesconi2014uvm,tavares2025case}, while ensuring accurate connections to the interfaces and protocol behaviours of the Design-Under-Test (DUT). As a result, verification testbench codebase typically grows to 4 or 5 times the size of the original RTL codes, imposing a substantial engineering burden.
% Among the verification execution loop, two major challenges significantly hinder efficiency. Initially, the construction of UVM-based verification environments. The Universal Verification Methodology (UVM) is widely adopted in industrial-scale verification for its support of modular, layered testbench architectures and reusable verification components. For instance, via sequencers, stimulus generation has been separated, which allows for reuse across different designs. However, this modularity comes at the cost of significant setup complexity. Engineers must manually instantiate and configure multiple hierarchical components according to SystemVerilog class conventions and UVM library specifications, while ensuring precise connectivity with the DUT’s interfaces and protocol behavior. Consequently, the verification environment codebase typically reaches 4 to 5 times the size of the original RTL, imposing a heavy development burden.

The second challenge lies in testcase supplement, which involves generating effective test cases to achieve comprehensive function coverage~\cite{pavithran2017uvm,dharani2023design,kumar2024robust}. This requires iterative refinement of test cases, including the careful generation of scenarios to capture boundary conditions, rare event sequences, and intricate state transitions. Coverage achieved, defined as the fulfilment of all user-defined function points, becomes increasingly more challenging as design complexity grows~\cite{salem2013modified,wu2023uvm,reddy2024uvm}. The manual identification and supplement of these scenarios into testbench are both costly and prone to human error. 

% Subsequently, the generation of high-quality stimuli for functional coverage closure presents a second major obstacle. This process involves iterative refinement of the test stimuli, including the creation of scenarios that capture boundary conditions, rare event sequences, and intricate state transitions. Achieving coverage closure, that is, the state in which all user-defined functional coverage goals have been met, becomes increasingly difficult as the complexity of the DUT and the state space grow. Manually identifying new scenarios and supplementing the testbench with directed stimuli is both time-intensive and error-prone.

Recent advances in large language models (LLMs) have opened new opportunities for automated assistance in hardware design and verification~\cite{thakur2023verigen,blocklove2023chip,tsai2023rtlfixer,ahmad2023fixing,fu2023llm4sechw,delorenzo2024make,liu2024domain,yao2024hdldebugger,fang2024assertllm,yao2024location,zhou2025insights}. Prior work has made promising strides: MEIC~\cite{xu2024meic} demonstrated the feasibility of employing LLMs in verification. However, it neither adopts a UVM-based verification flow, thus remaining disconnected from industry-standard methodologies, nor scales beyond small designs (typically fewer than 150 lines of RTL), limiting its applicability to realistic industrial settings. UVLLM~\cite{hu2024uvllm} improved upon this by integrating LLMs into a UVM-compatible workflow, demonstrating that LLMs can assist in generating UVM components. Nevertheless, the overall verification testbench still demands manual construction. Furthermore, both approaches rely on randomised stimulus generation without targeted coverage refinement or feedback-driven optimisation, leading to limited code and function coverage achievements and reduced verification efficiency.

In this paper, we present an LLM-aided UVM Machine (\textbf{\name}), the first systematic framework to realise an automated, LLM-driven function verification. 
By integrating domain-knowledge-guided prompt engineering with syntactic rule constraints~\cite{white2023prompt,liu2023verilogeval,chang2023chipgpt,sahoo2024systematic}, {\name} automatically generates UVM-based verification testbench, and iteratively refines test stimuli based on coverage feedback. This approach not only lowers the barrier to adopting UVM-based verification by significantly reducing the need for human involvement, but also turns a promising concept into a practical and scalable solution, demonstrating improved verification outcomes (e.g., more than 20\% improvement over the the state-of-the-art solutions) and increased efficiency (e.g., reducing setup time by more than 15×). 
% \begin{figure*}[htbp]
%     \centering
%     \includegraphics[trim=10mm 00mm 10mm 00mm, clip, width=1\textwidth]{./graphis/draft1.2/fig2.pdf} % 可根据需求调整宽度，如 0.9\textwidth
%     \caption{
%     % The UVMatic Framework Overview: The process begins with the DUT and the Specification (Spec.), the top level configuration. Initially, the DUT is pre - processed to eliminate syntax and focused timing - related errors (Step \textcircled{1}). Subsequently, the pre - processed code is tested under a UVM testbench (Step \textcircled{2}), and the log is then post - processed to extract relevant data (Step \textcircled{3}), which the debug agents use to generate candidate patches (Step \textcircled{4}). These codes and their pass rates are archived in the Repository (Repo.) and Register (Reg.) for future iterations.
%     The UVMatic Framework Overview: The process starts with the Specification (Spec.),and the configuration and DUT. Initally, extract the specified DUT module from the configuration (Step \textcircled{1}).  Subsequently,  use LLMs (Generation Agents) and scripts to design a UVM testbench from the bottom - up  (Step \textcircled{2}). Then, use VCS to complete the verification process of the testbench(Step \textcircled{3}). If it fails, use LLMs (Repair Agents) for error repair (Step \textcircled{4}).  Finally, use LLMs (Optimization Agents) to improve coverage that has passed the simulation (Step \textcircled{5}).  Both Step \textcircled{4} and Step \textcircled{5} are iterative, limited by the max iteration count.}
%     \label{fig:UVMatic_framework} % 便于后续引用
% \end{figure*}

The main contributions of this paper are:

\begin{itemize}
% \item \textbf{A fully automated framework for UVM-based verification environment generation:} {\name} leverages LLMs with domain-guided prompting and template (\textcolor{red}{template has not yet defined upto now, so reader won't know what it is}) to generate industrial-level, UVM-compliant verification environments, significantly reducing manual overhead.
\item \textbf{An automated framework for UVM testbench generation:} {\name} employs LLMs guided by domain-specific strategies to produce industrial-grade, UVM testbench, significantly alleviating manual development overhead.
\item \textbf{Automated stimuli generation with an iterative refinement:} {\name} iteratively improves function coverage by analysing collected coverage data and supplementing test stimuli, thereby accelerating coverage achieved.
\item \textbf{End-to-end integration into practical verification workflows:} {\name} supports the complete functional verification, demonstrating its scalability and adaptability across diverse and real RTL designs and protocols.
\item \textbf{Open-source release of the {\name} framework and benchmark:} we release the complete {\name} framework and all component within it to support wider adoption. In addition, we publish our evaluation benchmark comprising 10 real RTL designs (ranging from 400 to 1,600 lines of code) for cross-comparison and future research at \url{https://anonymous.4open.science/r/UVM_Machine-7806/}.
% \item \textbf{Demonstrated performance improvement:} \textcolor{red}{less important contribution.} 
% \name\ achieves a 90\% success rate in UVM environment generation across diverse modules, delivering a 100× speedup in testbench construction against manual development. Its testbenches reach 76.4\% initial functional coverage, improving to 98.1\% after an average of 3.2 optimization iterations, over 9× higher than MEIC and other SOTA baselines.
\end{itemize}

The rest of this paper is organised as follows. Section~\ref{sc:app} introduces the overall architecture of \name. Section~\ref{sc:det} then details the key techniques in each workflow and their rationales. 
Section~\ref{sc:eva} assesses our work based on the five proposed research questions.
% , followed by the related work given in Section~\ref{sc:re}. 
Section~\ref{sc:con} concludes the paper and outlines future research directions.

% \zhenote{Contributions are: 
% 1) Fully automated generation of UVM framework.\\
% 2) Stimuli generations.\\ 
% 3) Demonstration of integration of our medology into IC development flow\\ 
% 4) Open-source framework\\
% 5) Open-source benchmark\\ 
% 6) Comprehensive evalution \\ 
% }

\section{\name: An Overview}
\label{sc:app}
% \name\ focuses on the verification stage and aims to shorten the time required to complete the verification execution loop through a collaborative workflow that integrates LLMs and templates. 

\begin{figure*}[t]
    \centering
    \includegraphics[trim=00mm 00mm 00mm 00mm, clip, width=0.95\textwidth]{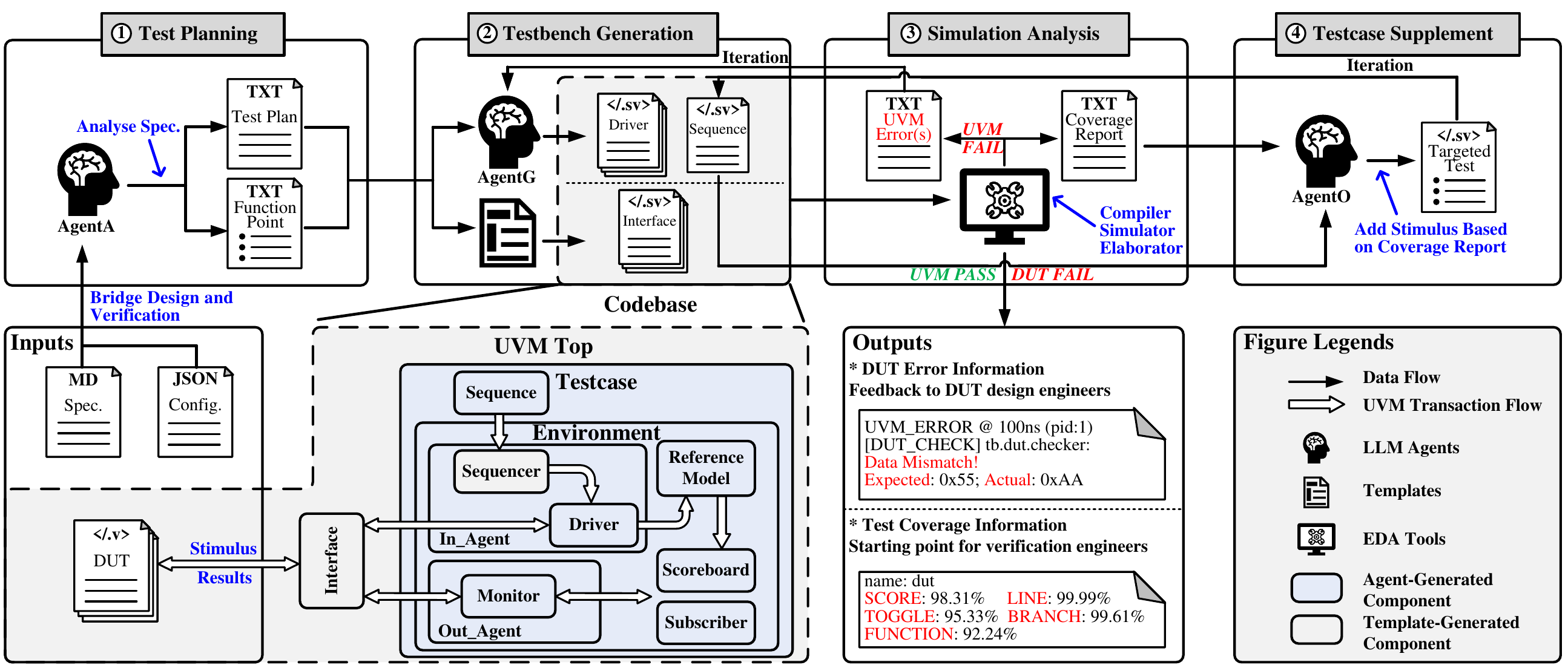} % 可根据需求调整宽度，如 0.9\textwidth
    \caption{
    % \textcolor{red}{Xinwei: All figure and table captions should follow the style of Fig. 1’s. Don’t repeat content that has already been covered in the text.}
    Overview of the {\name} Framework, which integrates UVM with LLM agents to automate the IC verification workflow. The framework includes Analysis Agent (AgentA) for test planning, Generation Agent (AgentG) for automatic testbench creation and error-driven regeneration, and Optimisation Agent (AgentO) for iterative testcase supplement based on coverage analysis. 
    % Analysis Agents (AgentA) assist in test planning by generating test plans and functional points from specifications. Generation Agents (AgentG) automatically create testbench components like Driver and Sequence. After test execution via EDA tools, UVM-related errors are sent to Generation Agents (AgentG) for regeneration while coverage data are analyzed by Optimization Agents (AgentO) to supplement testcases iteratively. 
    }
    \vspace{-10pt}
    \label{fig:flow} % 便于后续引用
\end{figure*}

% Aiming to accelerate the verification execution loop, \name\ utilises task-specific LLM agents with templates within a standardised UVM tool-chain as shown in Fig.~ \ref{fig:flow}. It comprises three specialised agents, one each for analysis, generation and optimisation, and a library of reusable templates that ensures generation accuracy and reduces LLM usage. \name\ produces the verification results as an error information report and a coverage report, and assumes the following inputs:
% \zhenote{Please highlight these three inputs are common in the human world!
% This means that we do not require any other stuffs.}

Aiming to accelerate the verification execution loop, \name\ utilises task-specific LLM agents with templates within a standardised UVM tool-chain as shown in Fig.~\ref{fig:flow}. It comprises three specialised agents, one each for analysis, generation and optimisation, and a library of reusable templates as predefined scripts that ensure generation accuracy and reduce LLM usage. \name\ produces the verification results as an error information report and a coverage report and assumes following three inputs:

% \name\ targets the verification stage, aiming to accelerate the verification execution loop through a collaborative workflow combining LLMs and templates. 
% As shown in \textbf{Figure \ref{fig:flow}}, the framework consists of a standardized verification toolchain, three categories of task-specific LLM agents (for analysis, generation, and optimization), and templates for the generation of partial UVM components.

% To ensure generalizability across different verification scenarios, \name\ standardizes its inputs as:
\begin{itemize}
    \item \textbf{Design specifications:} which define the intended functionality and behaviour of the DUT, written in Markdown format.
    \item \textbf{Configuration files:} in which the user specifies the target DUT and the reset state using a JSON file.
    \item \textbf{DUT:} the design entity implemented in RTL code.
\end{itemize}

These inputs are already expected in existing industrial verification and align with the standard workflow followed by verification engineers, enabling \name\ to integrate seamlessly without the need for additional formats or tool dependencies. To maximise accuracy while keeping token usage predictable, each LLM agent is fine-tuned and guided by domain-specific prompts. The overall flow is developed through the following iterative processes:
\begin{itemize}
    \item \textbf{Test Planning.} The Analysis Agent (AgentA) examines the input design specification to identify function points and constructs a coverage-driven testcase plan. 
\item \textbf{Testbench Generation.} Leveraging the identified test plan, the Generation Agent (AgentG) and the pre-validated templates jointly produce a UVM-based verification testbench. 
% The Generation Agent synthesizes key components such as drivers, monitors, and reference models, while templates generate the remaining modules (see Fig.~\ref{fig:flow}). 
\item \textbf{Simulation Analysis.} 
% The generated UVM testbench is executed using commercial EDA tools. 
Any errors in the generated UVM testbench are captured during simulation and used to refine the verification testbench, while coverage data are also collected for analysis and supplement. 
\item \textbf{Testcase Supplement.} If coverage remains insufficient, the Optimization Agents (AgentO) create additional targeted sequences to improve stimulus diversity, iteratively refining the testbench until coverage goals or user-defined iterations are reached. 
\end{itemize}

\parlabel{Modularity and flexibility.} 
% Notably, each stage in \name\ is connected via standardized interfaces, enabling smooth upgrades and flexible extensions. For example, by maintaining a consistent configuration format, users can replace the LLM model with a domain-specific model or switch to an alternative verification toolchain. This flexibility allows \name\ to adapt to diverse verification scenarios and evolving technologies. As a UVM-based automated verification framework, \name\ is also compatible with a wide range of existing optimization techniques to further enhance UVM verification efficiency. In this work, we validate its effectiveness using the Deepseek R1 model (see Section \ref{sc:eva} for details).
To allow \name\ to evolve with future verification techniques, all artefacts are exchanged via well-defined formats, enabling designers to swap in a domain-specific LLM, replace an EDA tool, or insert additional processes without impacting the validity of the workflow.

% Each stage in \name\ is interconnected through standardized interfaces, allowing seamless updates and extensions. For instance, with consistent configuration formats, users can swap the LLM model with a domain-specific model. This flexibility enables \name\ to accommodate various verification scenarios and evolving technologies. As an automated UVM-based framework, \name\ is also compatible with a broad range of optimization techniques to further improve verification efficiency. 

% \textcolor{red}{this statement is related to implementation, and shouldn't be here}Its effectiveness is validated using the Deepseek R1 model (see Section \ref{sc:eva} for details).

% \section{The Framework Pipeline}
% \label{sc:det}
%  \textcolor{purple}{["Not updated yet, needs more thought."]

%  [The later part has been revised.]
%  }

\section{The Framework Pipeline}
\label{sc:det}
% \textcolor{red}{This section needs to be improved significantly—learn from the MIEC Section 3. Maybe Yuchen and Ke can help. }
 % We first introduce testcase mapping strategies (Section \ref{subsc:input}); Section \ref{subsc:gen} details the UVM environment generation process; Section \ref{subsc:err} describes testcase analysis algorithm, including error handling; and Section \ref{subsc:cov} presents testcase supplementation to improve coverage.
We first present test planning from input specifications (Section~\ref{subsc:input}) and UVM testbench generation (Section~\ref{subsc:gen}), both serving as the foundation of the framework. We then describe simulation analysis with error handling (Section~\ref{subsc:err}) and coverage-driven supplementation (Section~\ref{subsc:cov}) to refine the UVM testbench and improve verification quality. Finally, we introduce the benchmarking setup used to assess framework performance (Section~\ref{subsc:ben}).
% \zhenote{This paragraph means nothing. But, it seems that we have introduce the main functionalities of the modules in previous section?}
% \textcolor{purple}{The functionalities truly have been described in sec.II. But as the starting paragraph, it's better to show the structure flow (just as overview)?}

% \subsection{\textcolor{purple}{Template and Agents Integration Strategies}}
% \subsection{\textcolor{purple}{Prompt-Driven Agent Collaboration and Template Use - step1(AA) step2(GA+Temp) step4(OA)}}
% \label{subsc:agent}
% Recent work such as MEIC and AssertSolver demonstrates that large language models (LLMs), when combined with domain knowledge and system-level prompting, can significantly improve performance on specialized tasks. Inspired by this, we fine-tune LLMs using instructions that emulate the workflow of IC verification engineers, enabling task-specific reasoning and automation.

% \textbf{Prompting techniques.} We categorize agents into three distinct roles, each aligned with a specific step in the verification execution loop:

% \textcolor{purple}{As shown in Figure 3, template-based generation achieves 100\% correctness, compared to 92\% using LLMs alone.}
% \begin{figure}[htbp]
%     \centering
%     \includegraphics[trim=20mm 50mm 20mm 50mm, clip, width=0.4\textwidth]{./graphis/draft1.1/test.pdf} % 可根据需求调整宽度，如 0.9\textwidth
%     \caption{Choose LLM or scripts}
%     \label{fig:LLM and scripts} % 便于后续引用
% \end{figure}

% \subsection{End-to-end Verification Workflow}

\begin{figure}[t]
    \centering
    \includegraphics[trim=00mm 00mm 00mm 00mm, clip, width=0.98\linewidth]{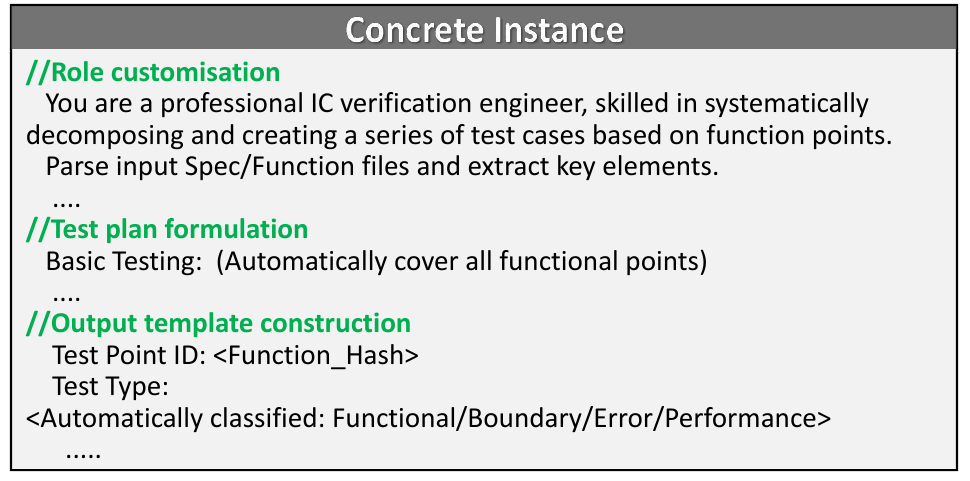} % 可根据需求调整宽度，如 0.9\textwidth
    \caption{Prompt Instructions for AgentA.}
    \vspace{-10pt}
    \label{fig:Any_prompt} 
\end{figure}

\subsection{Test planning}
\label{subsc:input}
% In IC design, while the design process is to implement functional behaviour,  such as enabling a calculator to perform basic arithmetic, the verification process is to ensure that all functional scenarios, including edge cases like division by zero, are correctly handled. This often entails identifying both expected and unexpected design behaviours.
% The cornerstone of effective verification lies in accurately interpreting the design specification and extracting function points that must be tested. This process typically requires significant domain expertise, as engineers must translate abstract design intents into concrete test objectives. In \name\, we aim to automate this step using an LLM-based analysis agent.

% However, general-purpose LLMs often struggle with complex specification reasoning due to their lack of structured memory and tendency to hallucinate. They may overlook critical corner cases or misinterpret loosely defined requirements. To address this, \name\ breaks down the analysis task into a structured reasoning pipeline that mimics the analytical behaviour of experienced verification engineers.
In IC design, the design process focuses on implementing functional behaviour, while functional verification ensures all possible scenarios, including edge cases, are handled correctly. Effective verification relies on interpreting the design specification and identifying function points for testing, a task requiring domain expertise. In \name, we automate this process with  LLM-based agents, AgentA. However, general-purpose LLMs often struggle with complex specifications due to their lack of structured memory and tendency to hallucinate~\cite{perkovic2024hallucinations,reddy2024hallucinations,tonmoy2024comprehensive}, leading to missed edge cases. To address this, \name\ breaks down the analysis into a structured reasoning pipeline that mimics expert verification engineers.

As shown in Fig.~{\ref{fig:Any_prompt}}, our approach follows a three-step analysis flow, designed to isolate the high-level cognitive tasks involved in human-driven verification planning:

\begin{enumerate}

\item\textbf{Role customisation:}
% The LLM is first guided to identify the functional expectations stated in the specification, including signal dataflow, control dependencies, input/output interaction patterns, and state transitions. By extracting these components, the agent gains insight into the functionality of the design.
AgentA is first guided through prompts to act as an IC verification engineer, enabling it to identify the functional expectations stated in the specification, including signal dataflow, control dependencies, input/output interaction patterns, and state transitions. From these elements, the agent extracts specific functional points, leading to a deeper understanding of the design’s intended behavior.

\item\textbf{Test plan formulation:}
% Once the workflow is abstracted, the agent derives testing strategies for each identified operation, specifying stimulus conditions, observability points, and coverage goals. For instance, testing an arithmetic unit involves exercising full operand ranges and signed/unsigned combinations under overflow and underflow scenarios.
based on the function points, AgentA is required to define testing strategies for each functional point, including stimulus conditions, observability points, and coverage goals. For example, testing an arithmetic unit involves exercising the full range of operands and all signed/unsigned combinations under both overflow and underflow scenarios.

\item\textbf{Output template construction:}
% To ensure consistency, analysis outputs are structured using a predefined template that records each functional point, its corresponding test strategy, and a draft testcase. This format streamlines downstream mapping to UVM components and supports human review and refinement.
to ensure consistency, all related files, including the test plan and functional points (stored in .txt format), follow a predefined template.  This template captures each functional point along with its associated test strategy and a draft test case, thereby enabling integration with UVM components and supporting manual inspection and refinement.

% \item\textbf{Key optimisation rule identification:}
% Finally, the agent identifies optional optimisations, such as stimulus grouping or corner-case prioritisation, based on heuristics derived from engineering practice. For instance, functionally equivalent branches may be merged to reduce testcase redundancy.

\end{enumerate}

This three-step process not only ensures that critical functional points are consistently extracted from the specification, but also improves LLM reliability by enforcing deterministic reasoning paths. The generated test plan serves as the foundation for subsequent verification stages, ensuring alignment between spec-driven objectives and testbench implementation.

\begin{figure}[t]
    \centering
    \includegraphics[trim=00mm 00mm 00mm 00mm, clip, width=0.92\linewidth]{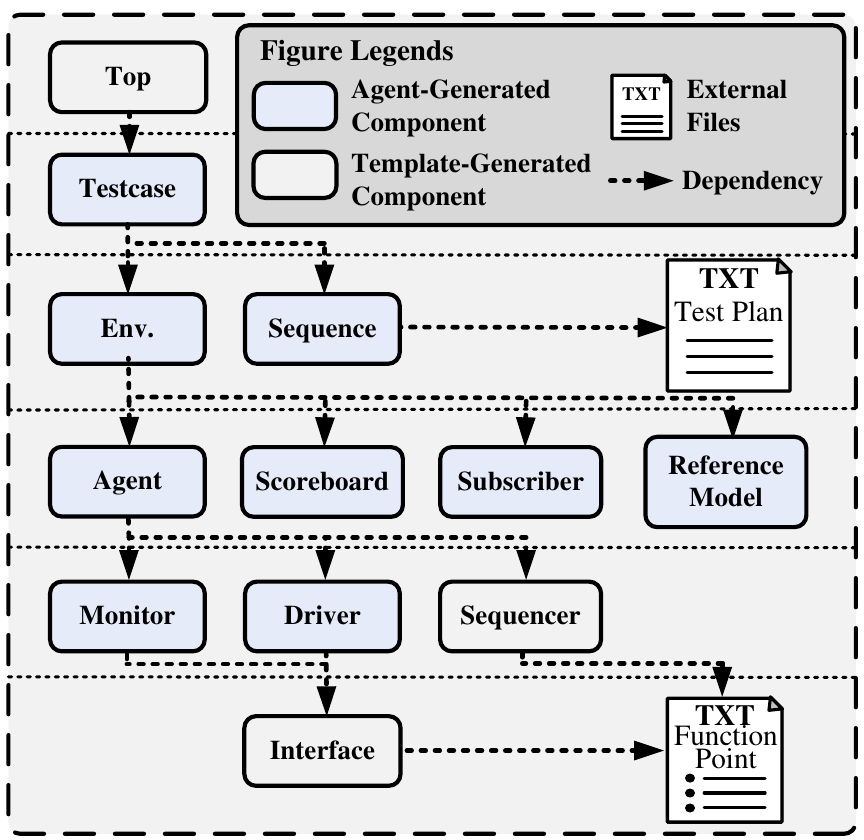}
    \caption{Dependency-Driven UVM Testbench Generation Workflow with Hierarchical Organization and External Dependencies.}
    \vspace{-10pt}
    \label{fig:Gen_flow} % 便于后续引用
\end{figure}

\subsection{Verification Testbench Generation}
\label{subsc:gen}
A robust verification testbench must not only support a broad range of stimuli and monitoring strategies but also reflect the structure and interaction logic of the underlying design. In UVM-based flows, the generation of testbenches is extremely time-consuming, largely due to component dependencies. UVM components are tightly coupled structurally and behaviourally. For instance, the Driver and Monitor are inseparable from the Interface: the Driver needs signal-level connectivity to drive transactions, and the Monitor must observe the same signals to collect functional responses. Likewise, an Agent encapsulates and coordinates multiple sub-components (Sequencer, Driver, Monitor), and cannot be constructed without knowing their details. Naïvely generating each module in isolation risks producing structurally invalid or semantically inconsistent testbenches.

\parlabel{Dependency-Driven UVM Testbench Flow.} To address this, \name\ employs a dependency-driven testbench generation flow as shown in Fig.~\ref{fig:Gen_flow} that sequences component creation based on interdependencies. The process begins with shared infrastructure (\eg, Interface), proceeds to leaf-level components (Driver, Monitor), then composites (Agents), and concludes with environment-level modules (Scoreboard, Top, Testcase). This ensures each component is generated with complete contextual awareness of its dependencies.

Once the flow is defined, component generation strategies must be selected. Analysis reveals that certain UVM modules exhibit deterministic and repetitive patterns, with variations primarily in signal names and data widths. These can be efficiently synthesised using template-based generation. In contrast, modules involving behavioural logic, such as transaction sequencing or protocol-specific checks, require semantic understanding and are better suited for LLM-based generation. Accordingly, \name\ employs a hybrid synthesis strategy (Fig.~\ref{fig:Gen_flow}), combining templates for structurally regular components with LLMs for functionally complex ones to optimise efficiency, correctness, and robustness.

\begin{figure}[t]
    \centering
    \includegraphics[trim=00mm 0mm 0mm 00mm, clip, width=0.98\linewidth]{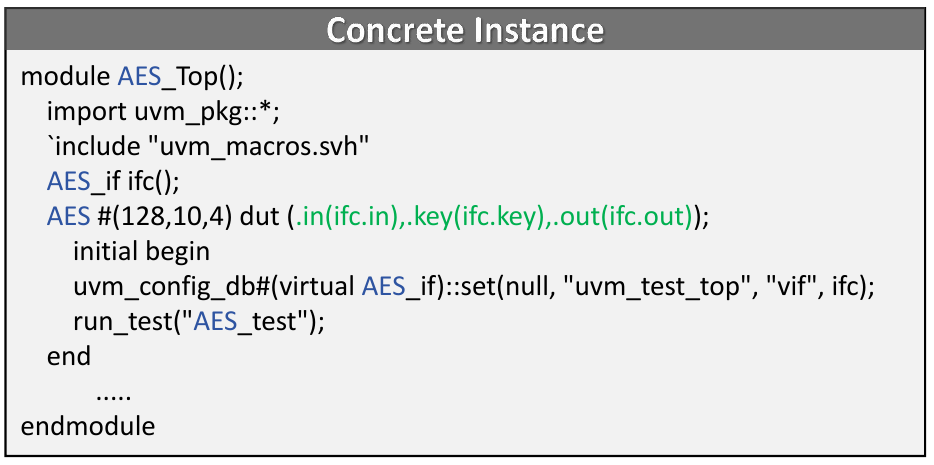} % 可根据需求调整宽度，如 0.9\textwidth
    \caption{The ``Top'' module Template in the Testbench. This template facilitates the integration of the testbench with the DUT. The DUT instance name (\textcolor{blue}{blue} text) and DUT ports (\textcolor{darkgreen}{green} text) are customizable parameters that allow the template to be configured for various verification scenarios.}
    \vspace{-10pt}
    \label{fig:Template} % 便于后续引用
\end{figure}

\begin{figure}[t]
    \centering
    \includegraphics[trim=0mm 0mm 00mm 00mm, clip, width=0.98\linewidth]{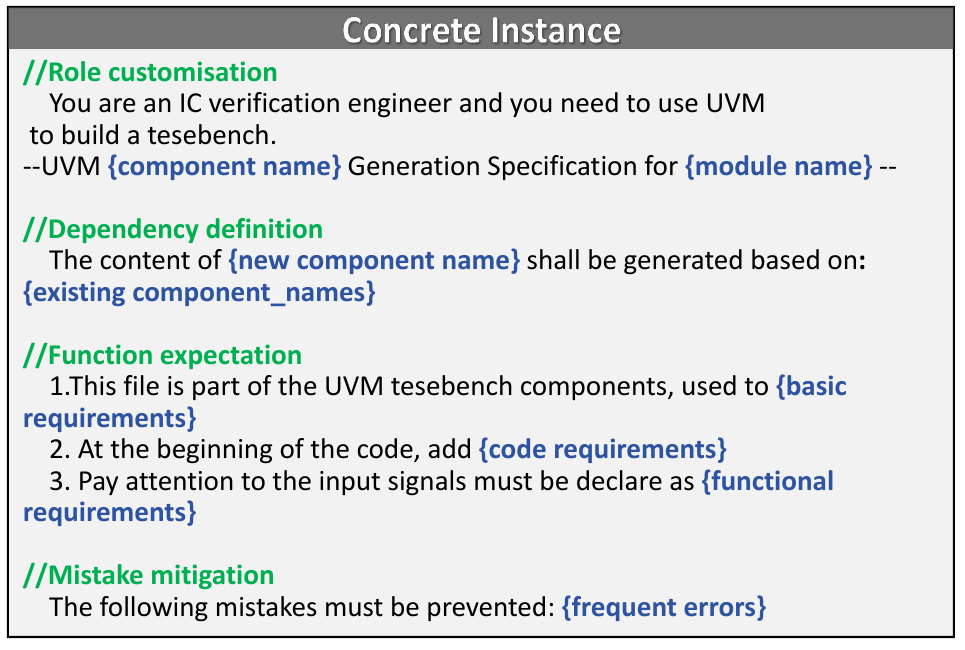} % 可根据需求调整宽度，如 0.9\textwidth
    \caption{Prompt Instructions for AgentG.}
    \vspace{-10pt}
    \label{fig:Gen_prompt} % 便于后续引用
\end{figure}

 \parlabel{Template-Based Generation.}
Templates are employed for modules that show high regularity across designs. This choice is motivated by three main factors: stability, speed, and reduced hallucination risk.
Interface modules follow near-identical formats, varying primarily in the number and naming of input/output ports.
Top modules serve as wrappers that instantiate all lower-level components and connect them via DUT ports; their structure is fixed and easily scriptable.
Sequencer modules exhibit a rigid control pattern that dispatches sequences to the Driver and rarely require complex logic.
Fig.~\ref{fig:Template} is an example generated by the top module using a template. After extracting the module name and port signal from the spec, the template will automatically fill in and generate the complete component.

 \parlabel{LLM-Based Generation.}
In contrast, components with complex behaviour, such as Driver, Monitor, and Scoreboard, require precise functional encoding tailored to specific testcase semantics and verification goals identified earlier. These modules necessitate adaptable code generation that aligns with both design behaviour and test intent. To support this, \name\ applies a structured prompting framework for each LLM generation task, as shown in Fig.~\ref{fig:Gen_prompt}. Each prompt is systematically constructed in four stages to guide the LLM in producing accurate, context-aware SystemVerilog code.

\begin{enumerate}

\item\textbf{Role customisation:}
% The agent receives information about the DUT, the verification objective, and the specific UVM component to be synthesized (for example, a Driver for a UART transmitter). This constrains the generation scope and enhances relevance.
we guide AgentG to act as an IC verification expert through prompt engineering, enabling it to more accurately understand the design functionality, verification objectives, and the required UVM components. This prompt-driven fine-tuning effectively constrains the relevance and scope of the generated content.

% \item\textbf{ Define component dependencies:}
\item\textbf{Dependency definition:}
dependency information is provided to ensure consistency with related components. For example, a Monitor must reference signal definitions from the Interface and transaction structures from the Sequence Item.

% \item\textbf{Describe Functional Role and Behaviour Expectations:}
\item\textbf{Function expectation:}
the prompt outlines the component’s functional responsibilities, such as capturing rising-edge events, introducing random timing variations, or detecting protocol violations. It also includes guidance on recommended coding patterns, such as TLM port usage.

% \item\textbf{List Common Pitfalls to Avoid:}
\item\textbf{Mistake mitigation:}
% Frequent mistakes are highlighted, including missing reset logic, inconsistent transaction identifiers, or incorrect signal directions. These constraints mitigate hallucination and promote domain-specific accuracy.
typical mistakes are outlined in the prompt to reduce the impact of hallucinations caused by AgentG, such as missing reset logic, inconsistent transaction identifiers, or incorrect signal directions.

\end{enumerate}

\begin{algorithm}[t]
{
\small
\SetAlgoLined
% $\vartriangleright$ \text{\texttt{Pre-processor}}\\
\KwIn{Verification log $V_{log}$}
\KwOut{repaired Uvm\_Component $UC_{True}$}
\SetKwFunction{FMain} {\normalfont Fixed\_Err}
    \SetKwProg{Fn}{Function}{:}{}
    \Fn{\FMain{\normalfont $V_{log}$}}{
    {
 	\emph{$/*\ Err:\ Error\ message\ of\ component\ */$}\\
        \emph{$/*\ GeneAgentMap=\newline
        \{\newline
        ``driver'' : ``G\_driver\_agent'', \newline
       \ ``monitor'': ``G\_monitor\_agent'',\newline
       ...\newline
        % \#\#\ For\ all\ components\ in\ testbench\newline 
        \} */$}\\

        \For{$phase$ $\in$ $phases$}
        {
        	\If{ErrInPhase($V_{log}$, $phase$)}
        	{
                $PhaseErr$ = GetErr($V_{log}$, $phase$);
                
                \For{$UC$,$Err$ $\in$ $PhaseErr_{i}$}
                {
                	$AgentID$ = GeneAgentMap[$UC$];\\
                	$UC_{True}$ = CallAgent($AgentID$,\ $Err$);
                }
            }
        }
        \textbf{return}\ $UC_{True}$\\
} }
\textbf{End Function}   
}
\caption{Repair Mechanism.}
\label{algo:repair}
\end{algorithm}

\setlength{\textfloatsep}{6pt}

Together, these four stages emulate the thought process of skilled verification engineers, ensuring the generated code is functionally valid and integrable into the overall testbench architecture.

By integrating dependency-driven planning, template-based generation for structural components, and guided LLM synthesis for behavioural modules, \name\ delivers a scalable and automation-friendly workflow for UVM testbench construction. This division of labour not only accelerates testbench generation but also improves modular correctness and robustness in industrial-scale projects.

\subsection{Simulation Analysis}
\label{subsc:err}

Once the UVM testbench is assembled, the DUT is simulated using Synopsys VCS to verify functionality and assess coverage. VCS compiles the DUT and testbench, executes generated testcases, and collects reports on function coverage, code coverage, and runtime logs, which are essential for evaluating testbench quality and identifying verification gaps. If design bugs are found, \name\ provides detailed diagnostics to assist in debugging. 

\parlabel{Repair Mechanism}. However, due to the generative nature of our framework, initial outputs from LLMs or templates may contain errors, such as syntax issues, interface mismatches, or improper sequencing, particularly when LLMs receive ambiguous context or insufficient dependency information. Inspired by prior work such as MEIC, which demonstrated the value of iterative LLM refinement through feedback, we introduce a repair mechanism into \name. As outlined in Algorithm \ref{algo:repair}, if simulation fails, the framework attempts to repair faulty components using LLMs by taking UVM error logs as structured prompts. These are then fed back into the AgentG to drive a second (or third) round of regeneration, focusing on correcting specific structural or behavioural issues.
This iteration cycle is repeated iteratively until one of two termination conditions is met: (i) the simulation completes successfully, or (ii) a predefined maximum iteration is exhausted.

% Once the simulation passes, the framework proceeds to coverage analysis, which serves as a critical metric for validating the completeness of the verification testbench. Coverage analysis includes both code and function coverage, with a threshold (\ie, 90\%) set for either metric. If the coverage meets the threshold, a final coverage report is generated. 
% Otherwise, if coverage remains insufficient, the coverage report is forwarded to the Optimization Agent, which analyses the gaps and supplements the stimulus.

% Testcase supplement is a critical phase in function verification, with empirical data showing that coverage closure can consume up to 50\% of the total function verification time (see Fig.~\ref{fig:Intro}). To close coverage gaps efficiently, \name\ introduces the Optimization Agent, a target-specific LLM module responsible for augmenting the test suite based on simulation feedback. The Optimization Agent addresses this bottleneck using a structured prompt flow, as illustrated in Fig.~\ref{fig:O_prompt}, guiding the LLM through five reasoning steps informed by engineering practices. This agent analyses coverage reports and generates new stimuli targeting uncovered function points, significantly accelerating convergence toward coverage goals.

\begin{figure}[t]
    \centering
    \includegraphics[trim=00mm 00mm 00mm 00mm, clip, width=0.98\linewidth]{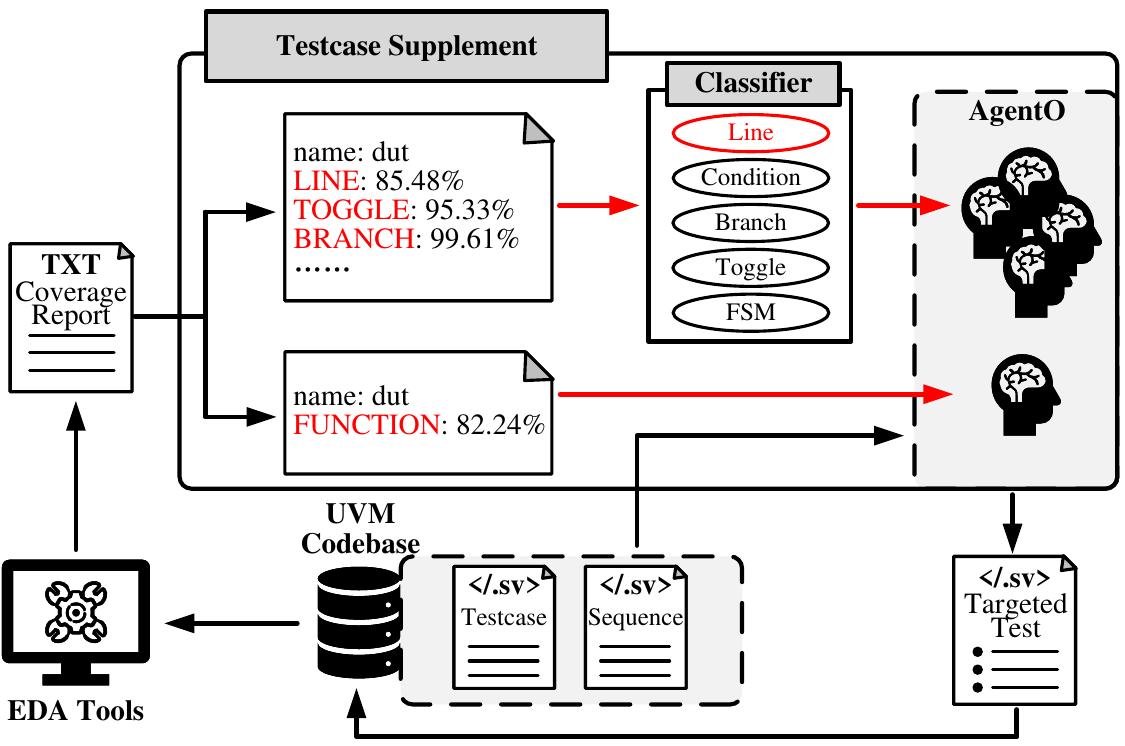} % 可根据需求调整宽度，如 0.9\textwidth
    \caption{Testcase Supplement Workflow with Coverage Analysis. }
    \vspace{-5pt}
    \label{fig:supple} % 便于后续引用
\end{figure}

\begin{figure}[t]
    \centering
    \includegraphics[trim=00mm 00mm 00mm 00mm, clip, width=0.98\linewidth]{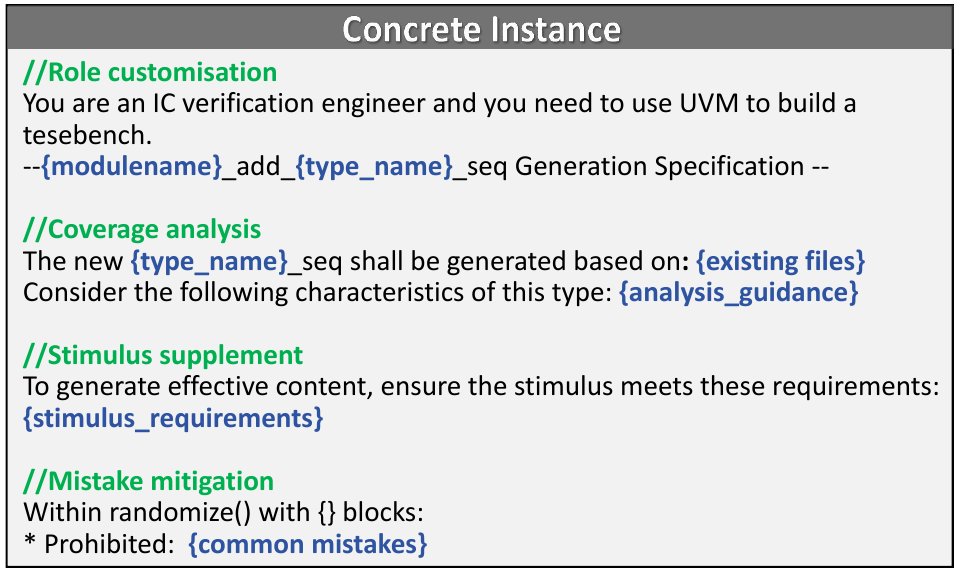} % 可根据需求调整宽度，如 0.9\textwidth
    \caption{Prompt Instructions for AgentO.}
    \vspace{-5pt}
    \label{fig:O_prompt} % 便于后续引用
\end{figure}

\subsection{Testcase Supplement}
\label{subsc:cov}

Ensuring the correctness of a testbench is only one part of the verification process—it does not by itself ensure completeness. In hardware verification, both code and functional coverage are essential indicators, and low or plateaued coverage often suggests that the applied stimuli are not effectively exercising key parts of the DUT. To tackle this issue, we introduce a testcase optimisation mechanism that enhances stimulus generation through sequence refinement, focusing on the protocol-level behaviours defined within UVM.

As shown in Fig.~\ref{fig:supple}, testcase supplement phase in \name\ analyzes coverage gaps, categorizing them by logic region, transaction stage, or stimulus dependency. Deficient types are passed to AgentO, which revises sequences to target these gaps, generating new testcases until satisfactory coverage is achieved. Precise prompting is crucial, as LLMs without guidance may repeat patterns or miss gaps, making goal-oriented context essential for effective coverage improvement.

To resolve this, we design a prompting instructions for each coverage type. As shown in Fig.~\ref{fig:O_prompt}, it includes four stages that collaboratively shape the LLM’s response:

\begin{enumerate}

% \item\textbf{Provide context and specify the target:} 
\item\textbf{Role customisation:} 
% We describe the DUT section, relevant protocol, and reference the specific coverage metric or type that remains unsatisfied.
We assign AgentO the role of an IC verification expert, equipped with sufficient knowledge to generate and optimize the UVM testbench. 

% \item\textbf{Provide analysis guidance for this type:} 
\item\textbf{Coverage analysis:} 
% We also supply the DUT section, relevant protocols, and reference the specific coverage metrics or types that remain unsatisfied to assist in the process.
We also provide the DUT section, relevant protocols, and reference the specific coverage metrics or types that remain unsatisfied. For the test points that do not satisfy the coverage criteria, we offer explanations to facilitate the process.

% We explain why the coverage is low for this type, \eg, branch not taken due to fixed operand values or lack of back-to-back transactions.

% \item\textbf{Specify requirements for this type:} 
\item\textbf{Stimulus supplement:} 
% This includes specific constraints, such as generating sequences with varying payload sizes, or triggering edge-case interrupts. 
This part includes specific constraints, such as generating sequences with varying payload sizes or triggering edge-case interrupts. These information are used to assist the LLM in supplementing missing stimuli.
% or testing error-handling logic.

\item\textbf{Mistake mitigation:} 
% To reduce LLM hallucination and increase signal-to-noise ratio, we explicitly forbid certain patterns, such as reusing sequences already covered or producing syntactically invalid phase orders.
To reduce LLM hallucination during stimulus supplementation, we explicitly forbid certain potentially problematic stimulus sequences in the prompt, including those with syntactic errors or functionally invalid behaviors.

\end{enumerate}
% By equipping the LLM with type-aware prompting and data-driven optimisation objectives, our framework improves testbench effectiveness in an automated, feedback-guided manner. Compared to static generation, this iterative supplement mechanism leads to substantial coverage gain with minimal human effort.
 \parlabel{Optimisation mechanism.} The refinement loop is designed to be fault-tolerant. If the updated testbench fails simulation, \name\ reverts to the last valid version. The loop continues until either the target coverage is achieved or the number of optimization iterations exceeds a user-defined maximum.
This closed-loop, coverage-driven refinement strategy mirrors expert workflows while dramatically reducing manual effort, resulting in faster, more reliable verification cycles.

\subsection{Benchmark}
\label{subsc:ben}
\begin{table*}[t]
    % \sffamily
    \small
    \belowrulesep=0pt
    \aboverulesep=0pt
    \centering
        \caption{Benchmark designs used for evaluation, including a variety of commonly implemented hardware modules such as cryptographic cores (AES, SHA256, SM4), arithmetic units (ALU), memory and communication controllers (DFI, SDRAM, SPI, UART), and a Huffman encoder.}
    \resizebox{0.9\linewidth}{!}{%
    \begin{tabular}{c|l|c|c }
    % {p{0.7cm} p{3.4cm}<{\centering} p{6.9cm}<{\centering}} 
    \toprule
     \textbf{Design} & \textbf{Description} & \textbf{\makecell{Module\\ Counts}}&\textbf{Line Counts} \\

    \midrule
         AES& Encrypts a 128-bit plaintext input using a configurable key (128, 192, or 256 bits) and outputs 128-bit ciphertext.  & 8 & 684 \\
         \midrule
         ALU& A 32-bit  unit that performs FP arithmetic, logical operations (OR, AND, XOR), shifts, FP to INT conversion, and complement.  & 7 & 409 \\\midrule
         DFI& A lightweight DDR3 memory controller that efficiently manages the interface between a system and DDR3 memory devices.  & 2 & 430 \\\midrule
         \multirow{2}{0.65cm}{HUF}& A synchronous Huffman encoding system that compresses 4-bit input data symbols by calculating their frequencies,  & \multirow{2}{0.35cm}{\textbf{10}} & \multirow{2}{0.6cm}{\textbf{1,572}} \\
         & generating Huffman codes, and outputting a compressed serial bitstream.&&\\\midrule
         SDRAM& A controller for synchronous dynamic random-access memory (SDRAM) with a Wishbone bus interface.  & 1 & 604 \\\midrule
         SHA256& A cryptographic unit that computes a 256-bit hash value from a 24-bit input message using the SHA-256 algorithm.   & 3 & \textbf{1,412} \\\midrule
         \multirow{2}{0.65cm}{SM4} & A synchronous implementation of the SM4 encryption/decryption algorithm, supporting both modes through a 128-bit key  & \multirow{2}{0.2cm}{\textbf{9}} & \multirow{2}{0.6cm}{\textbf{1,142}} \\
         & and data interface.&&\\\midrule
         SPI& A lightweight SPI controller supporting Master mode with FIFO-based data transfer and interrupt-driven operation.   & 1 & 839 \\\midrule
         UART& Facilitates serial communication between a host and peripherals, supporting configurable baud rates and stop bits.  & 3 & 639 \\
         
         \bottomrule
    \end{tabular}}
    \vspace{-6pt}
    \label{tab:ben}
\end{table*}

While simpler benchmarks such as RTLLM~\cite{lu2023rtllm} and Verilog-Eval~\cite{liu2023verilogeval}, which target designs with fewer than 200 lines of code, are useful for testing basic LLM capabilities in RTL design and verification, they do not accurately reflect the complexities of industrial scenarios. As such, they are not suitable for assessing the full scope and scalability of \name\ in real-world applications.

To evaluate the effectiveness of \name, we curated a diverse set of hardware design modules that collectively represent a wide spectrum of real-world verification challenges, as illustrated in Table~\ref{tab:ben}. These benchmarks span cryptographic cores (AES, SHA256, SM4), arithmetic units (ALU), memory controllers (DDR3, SDRAM), data compression (Huffman Encoder - HUF), and peripheral communication interfaces (SPI, UART). Each module presents unique structural and behavioral characteristics, ranging from highly sequential logic (\eg, SHA256, HUF) to pipelined arithmetic computation (\eg, ALU) and complex control flow (\eg, DDR3, SDRAM).

We deliberately selected modules with varying design sizes and complexities, as reflected by their line counts, to assess the scalability and adaptability of our approach. For instance, the AES and HUF modules involve intricate data-path logic and state machines, posing significant coverage challenges. Memory controllers like DDR3 and SDRAM demand rigorous protocol compliance, while modules such as UART and SPI, though smaller, require nuanced handling of asynchronous communication and timing constraints. 
% Together, this benchmark suite provides a comprehensive platform to evaluate how well an LLM-based system can generalize across different design paradigms, interface protocols, and verification scenarios.

\section{Evaluation}
\label{sc:eva}
% This section describes our experimental setup, research questions, evaluation metrics, and results to address the posed inquiries.

% \parlabel{Setup.} In our experiment, we employed LLM agents via the Deepseek API, with Deepseek-v3 as the default model. We set the temperature of the agent, which controls the LLM’s output randomness, to 0.3. Then we used our newly-proposed benchmark as the experimental object and adopted VCS as the simulation tool for our verification environment. Based on pre-experiment results, we set the error iteration times for Testcase Analysis to 2 and the optimization iteration times for Testcase Supplement to 2, since little improvement can be observed thereafter.
This section presents our experimental setup, evaluation metrics, research questions, and results to answer the questions.

% \parlabel{Setup.} In our experiment, we employed LLM agents via the Deepseek API, with Deepseek-v3 as the default model. We set the temperature of the agent, which controls the LLM’s output randomness, to 0.3. Then we used our benchmark as the experimental object and adopted VCS as the simulation tool for our verification testbench. Based on pre-experiment results, we set both the error iteration times and the optimization iteration times to 2, since little improvement can be observed thereafter.

\parlabel{Experimental setup.} In our experiment, we employed LLM agents via the Deepseek API, with Deepseek-v3 as the default model. 
We set the temperature of the agents, which controls the output randomness of the LLM, to 0.3. Then we used our benchmark for assessment and adopted VCS as the simulation tool for our testbench. 
Additional, we set the iteration times of the iteration mechanism in both Testcase Analysis and Testcase Supplement to 2, since little improvement can be observed based on pre-experiment results. 
Following the pass@5 evaluation standard commonly used in the LLM field, each component was generated five times per module to minimise experimental randomness.

\subsection{Evaluation Metrics}
To evaluate the performance of our work in correct and complete UVM-based testbench, we propose the following metrics:

\parlabel{Success Rate of Generation ($SR_{G}$):} 
% We take the UVM environment manually built by experienced engineers as the baseline. If the generated UVM Testbench can pass the vcs simulation process and align with the expected results, it is deemed correct; otherwise, it is labeled a generation failure.\textcolor{red}{Need to add formula}
This metric evaluates the correctness of the generated UVM testbenches by comparing them against a reference testbench manually constructed by experienced engineers. A generated testbench is considered correct if it passes the full VCS simulation flow and produces results that match the expected outputs defined by the reference testbench. Otherwise, it is categorized as a generation failure. This metric reflects the functional validity of the generation framework. 
\begin{equation} 
 SR_{G}= \frac{N_{\text{correct}}}{N_{\text{total}}} \times 100\% 
\end{equation}

\parlabel{Coverage:}  
% This serves as an index to quantify test completeness after the testbench passes the vcs simulation process, encompassing code coverage and functional coverage. The former, automatically generated by the simulation tool, measures the proportion of RTL code executed during simulation; the latter refers to the implementation proportion of functional scenarios realized by the RTL code versus the expected functional scenarios in the specification. Both indices are widely recognized in the hardware verification industry for quantifying verification completeness.
This metric quantifies the test completeness of a generated testbench after it has successfully passed the simulation phase. It is composed of two standard components:
\begin{itemize}
    \item Code Coverage ($C_{\text{code}}$): Automatically provided by simulation tools such as VCS, this measures the percentage of RTL code exercised by the testbench (\eg, line, branch, toggle coverage).
    \item Functional Coverage ($C_{\text{func}}$) : This represents how thoroughly the testbench exercises the RTL design with respect to the functional scenarios outlined in the design specification.
\end{itemize}
% The overall coverage score combines both metrics, which can be weighted depending on the verification goals:

% \begin{equation}
% \text{Coverage} = \alpha \cdot C_{\text{code}} + \beta \cdot C_{\text{func}} 
% \end{equation}
% $\alpha$ and $\beta$ are weighting factors such that $\alpha + \beta = 1$. Commonly, $\alpha = \beta = 0.5$, unless emphasis on one is desired.

\parlabel{Completion Time:} This denotes the time to complete the entire functional verification process depicted in Fig.~\ref{fig:Intro}. In our work, it corresponds to the time to execute the entire functional verification process under the premise that \name\ generates a correct Testbench.

\subsection{Research Questions}
We conducted experiments to evaluate \name\ with respect to five key research questions (RQs):

% \noindent \textbf{RQ1: How is the effectiveness of \name\ in generating correct UVM testbench?}
% This research tested the generation of 11 types of UVM components and also used three LLM models (\textcolor{red}{GPT4o, a closed-source model, and DS-v3 and Qwen, open-source models widely used in the LLM field with strong problem-solving abilities}) for generation as a comparison. To reduce experimental randomness, referring to the commonly used evaluation index pass@5 in the LLM field, we generated each component of each module 5 times.
% This research tested the generation of 11 types of UVM components with the use of various LLM models for generation as a comparison. To reduce experimental randomness, referring to the commonly used evaluation index pass@5 in the LLM field, we generated each component of each module 5 times.
\noindent \textbf{RQ1: How effective is \name\ in generating syntactically and semantically correct UVM testbenches?}
This research question evaluates the capability of \name\ to produce valid UVM testbenches that conform to standard structure and compile without errors.

% \noindent \textbf{RQ2: How does the verification completeness of \name\ compare with existing LLM-based verification work?}
% This research tested the coverage of UVM-based verification testbenches for 9 modules in the benchmark generated by \name\ and compared it with two instances (MEIC and UVLLM) of LLMs used in the IC verification.
% We measured the coverage of the testbench generated by \name\ across 9 modules and compared it with two instances (MEIC and UVLLM) of LLMs applied in function verification. Notably, since functional coverage relies on the setting of expected functional scenarios by humans, the function coverage in this experiment used the functional scenarios defined by experienced engineers when constructing the testbench as the expected scenarios.
\noindent \textbf{RQ2: How does the verification completeness achieved by \name\ compare to existing LLM-based verification approaches?}
This question investigates whether \name\ can generate testbenches that reach comparable or higher code and functional testcase coverage than other LLM-driven methods.

% \noindent \textbf{RQ3: How does the time for \name\ to complete the entire functional verification process compare with that of human experts?}
% We recorded the time required for \name\ to complete each process and the total time, comparing it with the time taken by experienced engineers to manually build a UVM environment. As shown in Figure 1, Testbench Generation and Testcase Supplement are bottlenecks, so we paid special attention to the time spent in these two stages.
% \noindent \textbf{RQ3: How does the end-to-end verification time using \name\ compare with that of expert engineers?}
\noindent \textbf{RQ3: How much of a performance gain in terms of efficiency can end-to-end verification using \name\ achieve compared to that of expert engineers?}
This question explores whether \name\ can reduce the total time needed to complete the functional verification process relative to manual effort.

% \noindent \textbf{RQ4: How much does the repair mechanism affect the generation success rate?}
% We focus on the generation success rate of UVM testbenches for 9 modules in each round under the action of \name\.To eliminate randomness, we conducted multiple tests for each module to correctly evaluate the necessity of the repair mechanism in Testcase Analysis (as described in Section {\ref{sc:det}}). It should be noted that “multiple tests” means starting \name\ multiple times—specifically, in one test, \name\ necessarily includes one or more generation attempts.
\noindent \textbf{RQ4: How does the repair mechanism influence the success rate of UVM testbench generation?}
This question examines whether incorporating automatic repair significantly improves the ability of \name\ to generate usable testbenches after initial failures.

% \noindent \textbf{RQ5: How much does the optimization mechanism improve the coverage?}
% We focus on the improvement of the coverage of UVM-based verification testbenches for each module by \name\ to demonstrate that the optimization mechanism is practical.
\noindent \textbf{RQ5: How does the optimisation mechanism in \name\ affect functional coverage improvement?}
This research question evaluates the effectiveness of the optimization strategy in increasing the achieved coverage metrics during simulation.

\subsection{Results and Discussions}
The results of our experiments are presented as follows, with each aligned to its respective research question.

\begin{figure*}[t]
    \centering
    \includegraphics[trim=00mm 00mm 00mm 00mm, clip, width=0.9\textwidth]{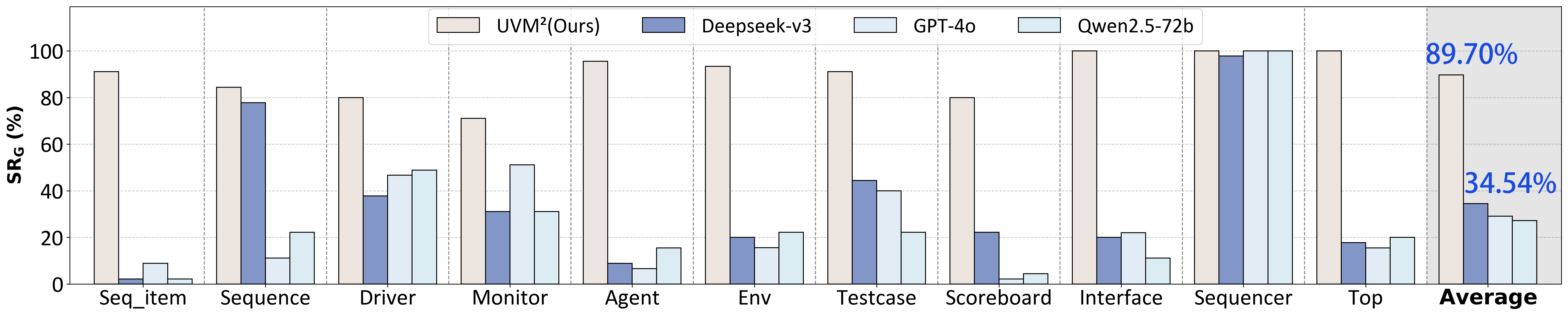} % 可根据需求调整宽度，如 0.9\textwidth
    \caption{$SR_G$ of 11 UVM components with \name\ against SOTA LLMs}
    \vspace{-5pt}
    \label{fig:ex1} % 便于后续引用
\end{figure*}

\begin{figure*}[t]
    \centering
    \includegraphics[trim=00mm 00mm 00mm 00mm, clip, width=0.92
\textwidth]{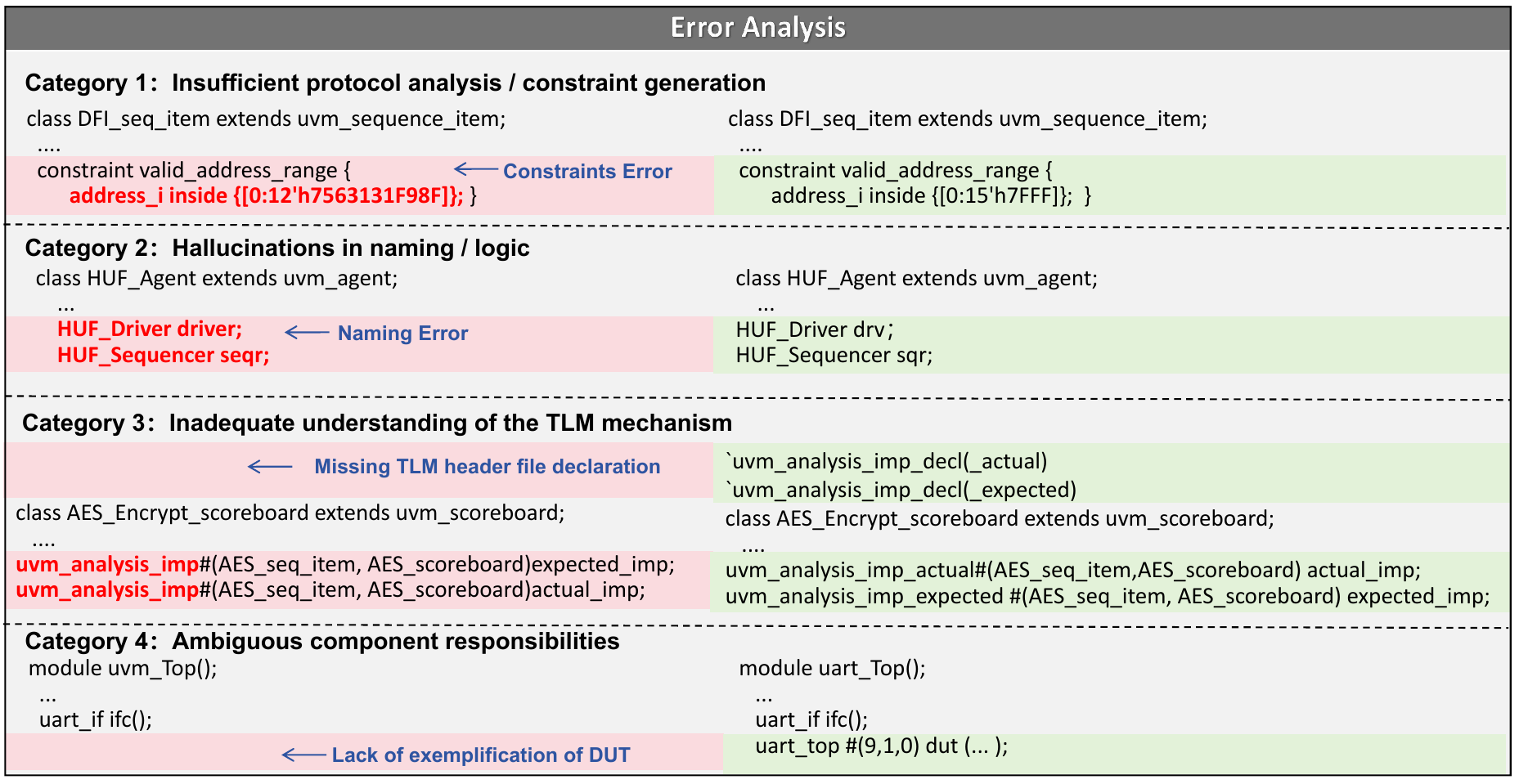} % 可根据需求调整宽度，如 0.9\textwidth
    \caption{Four categories of errors in LLM-generated UVM components and their corrections. 
    % Category 1 shows an incorrect address\_i constraint (left) and the corrected version (right). 
    % Category 2 highlights naming errors, such as HUF\_Driver driver and HUF\_Sequencer seqr (left) versus the corrected HUF\_Driver drv and HUF\_Sequencer sqr (right). 
    % Category 3 illustrates a missing TLM header file declaration (left) and the correct declarations (right). 
    % Category 4 shows the lack of DUT instantiation in the incorrect uvm\_Top module (left) and the correct uart\_Top module (right). 
    }
    \vspace{-5pt}
    \label{fig:error} % 便于后续引用
\end{figure*}

\begin{figure*}[t]
    \centering
    \includegraphics[trim=00mm 00mm 00mm 00mm, clip, width=0.92\textwidth]{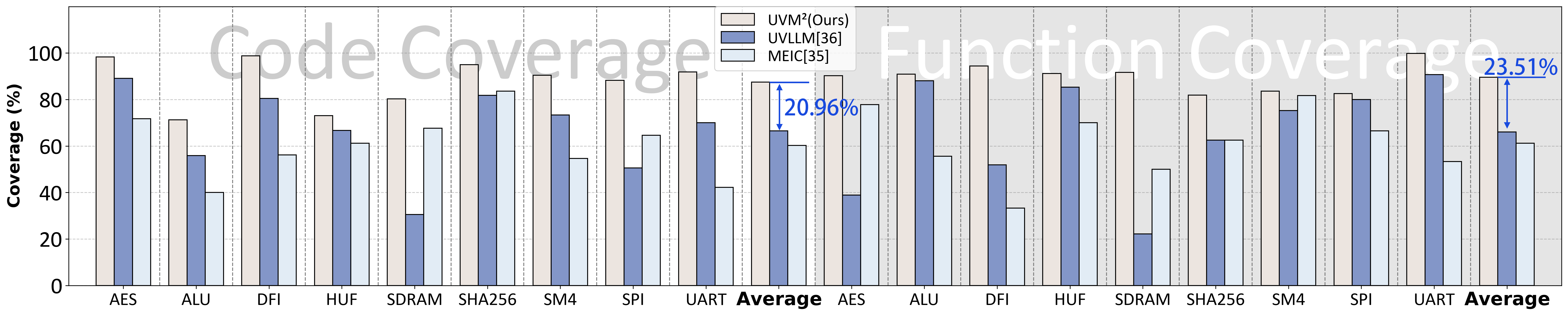} % 可根据需求调整宽度，如 0.9\textwidth
    \vspace{-5pt}
    \caption{Code Coverage and Function Coverage of \name, UVLLM, MEIC}
    \vspace{-15pt}
    \label{fig:ex2} % 便于后续引用
\end{figure*}

\noindent \textbf{Result 1:}
Fig.~{\ref{fig:ex1}} illustrates the $SR_G$ of  \name\ for various UVM components, achieving an average success rate of 89.70\%. Notably, the three components generated from the template technique achieve a 100\% success rate, validating the reliability of this approach. Among components generated by AgentG, agent, env, testcase, and seq\_item exceed 90\% $SR_G$. However, driver and monitor, two components requiring in-depth domain knowledge of timing information like handshakes, show lower success rates (approximately 80\%), reflecting the challenge of encoding complex designs without explicit guidance.

\name\ is also compared with three widely-used LLM baselines: GPT-4o (closed-source), Deepseek-v3, and Qwen2.5-72b (two open-source models widely adopted for their strong problem-solving capabilities). 
To ensure experimental rigor, each generated component was used to replace the corresponding module in a manually crafted UVM testbench developed by experienced engineers, while the remaining components were kept unchanged. This approach ensures a valid evaluation since individual components cannot be tested in isolation due to inherent interdependencies, as illustrated in Fig.~\ref{fig:Gen_flow}. For each generation task, the rest of the testbench was provided as reference context.
Among these baselines, Deepseek-v3 achieves the highest overall success rate at 34.54\%. 
In contrast, \name\ reaches a success rate that is 2.59× higher, surpassing all baselines across every component category. Notably, in the agent and seq\_item categories, where \name\ particularly excels, baseline performance is minimal, as agent success rates are 8.88\%, 6.66\%, and 15.55\%, and seq\_item success rates are 2.22\%, 8.89\%, and 2.22\% for Deepseek-v3, Qwen2.5-72b, and GPT-4o, respectively. This significant discrepancy underscores critical limitations in baseline capabilities.
% \zhenote{Explain below stuffs.}

Fig.~{\ref{fig:error}} provides concrete examples of such errors and their corrected versions, illustrating how LLM outputs deviate from UVM design norms without structured guidance.

\parlabel{Insufficient protocol analysis/constraint generation.} UVM requires precise sequence constraints (\eg, DFI protocol specifications). Legacy LLMs fail to parse protocols, generating invalid constraints (\eg, incorrect address ranges). This oversight can derail simulations entirely, as critical scenarios remain unaddressed, not merely reducing coverage but rendering verification incomplete and unreliable

\parlabel{Hallucinations in naming/logic.} LLMs invent invalid UVM class names or misapply methods. Naming like HUF\_Driver driver (as presented in Fig.~{\ref{fig:error}}) isn’t inherently wrong. However, in this UVM component, other parts are named as drv. This naming inconsistency leads to mismatches during instantiation, triggering runtime errors. Such lapses in naming coherence violate UVM’s structural norms, severely undermining the testbench’s reliability.

\parlabel{Inadequate understanding of TLM mechanism.} 
% LLMs often omit critical TLM declarations. Without proper uvm\_analysis\_imp\_decl (as missing in the incorrect code), subsequent method misuse like flawed uvm\_analysis\_imp instantiation follows. This two-fold issue—missing foundational declarations and misapplying methods—disrupts TLM-based communication, a cornerstone of UVM verification, and renders components non-functional.
LLMs often omit critical TLM declarations, such as uvm\_analysis\_imp\_decl. Without these foundational elements, subsequent method usage becomes flawed, including incorrect instantiation of uvm\_analysis\_imp. This combination of missing declarations and misapplied methods disrupts the TLM-based communication that underpins UVM verification, ultimately rendering components non-functional.

\parlabel{Ambiguous component responsibilities.} Each UVM module must adhere to its role. For instance, the top module’s duty includes proper DUT instantiation. Failing this (as in the incorrect example) violates the principle of clear - cut responsibilities. Such role - neglect doesn’t just cause minor issues but invalidates the entire verification setup, highlighting that overlapping or overlooked duties can collapse the UVM testbench’s effectiveness.invalidates testbench correctness.

These results demonstrate that relying solely on legacy LLMs for UVM testbench generation is infeasible due to domain knowledge gaps. By contrast, \name\ significantly enhances feasibility through targeted techniques (e.g., templates, iterative repair), achieving robust performance across both simple and complex components.

% \noindent \textbf{Result 2:}
% Fig.~{\ref{fig:ex2}} shows the coverage of \name\ and the other two control groups in each module. Specifically, the average code coverage of \name\ reaches 87.44\%, which is 20.96\% and 27.24\% higher than UVLLM and MEIC respectively. In terms of functional coverage, \name\ achieves 89.58\%, which is 23.51\% and 28.40\% higher than UVLLM and MEIC respectively. As shown in Fig.~{\ref{fig:ex2}}, the coverage of \name\ in the verification of each module is higher than that of the other two works. Among them, DDR3, with 430 lines of code and only two modules (elaborated in Table 1), achieves the highest code coverage of 98.31\% and functional coverage of 94.44\%. As the code complexity increases, the coverage of \name\ decreases. HUF, an RTL code with nearly 1,500 lines of code and containing ten modules, still has a code coverage of 73.07\% and a functional coverage of 91.21\%. However, the coverage of the results of the other two works varies among different modules, which is because the test stimulus generation of both of them is based on random stimulus generation, resulting in instability and incompleteness. This result indicates that \name\ significantly improves the test completeness in LLM-based verification. Meanwhile, \name\ also practices the theory of applying LLM to UVM-based verification, demonstrating that LLM-based testing does not have to rely solely on random stimuli.

\noindent \textbf{Result 2:}
We measured the coverage of the testbench generated by \name\ across 9 modules and compared it with two instances (MEIC and UVLLM) of LLMs applied in functional verification. Notably, since function coverage relies on the setting of expected functional scenarios by humans, the function coverage in this experiment used the functional scenarios defined by experienced engineers when constructing the testbench as the expected scenarios.

\begin{table*}[t]    
\sffamily    
\small    
\belowrulesep=0pt    
\aboverulesep=0pt    
\renewcommand{\arraystretch}{1}    
\centering     
\caption{Completion Time of \name\ against Human Verification Engineers. The time statistics for each stage incorporate multiple generation attempts as required. The ``Simulation'' time includes compile, elaboration, simulation and repair iterations for components.}     
\label{Table:runtime2}
     \begin{threeparttable}     
\resizebox{0.68\linewidth}{!}{%    
\begin{tabular}{c|  c  c c  c c|  c|c }    
% {c|  p{1.0cm}  p{1.2cm} p{1.2cm}  p{1.3cm} p{1.3cm}|  c|c }    
%p{1.2cm} p{1.2cm}| p{1.2cm} p{1.2cm}| p{1.2cm} p{1.2cm}| p{1.2cm} p{1.2cm}}    
\toprule    
\multirow{2}{*}{\centering\textbf{Design}} & \multicolumn{5}{c|}{\centering\textbf{\name}} &{\centering \textbf{Human}}&    \multirow{2}{*}{\centering\textbf{Speedup}} \\    
&   Planning& Generation&Simulation & Supplement& Total& Total& \\    
\midrule            
AES&  3.55min &12.52min &6.47min &5.83min &28.37min &16.22h &34.30x \\         
ALU&  3.43min &15.22min &5.28min &8.51min &32.44min &16.13h &29.83x \\         
DFI&  2.12min &11.53min &5.90min &5.41min &24.96min &16.15h &38.82x \\         
HUF&  25.45min &72.22min &40.85min &25.42min &163.94min &40.12h &14.68x \\         
SDRAM&  32.18min &80.22min &45.24min &25.35min &182.99min &39.34h &12.90x \\         
SHA256&  22.48min &48.43min &30.52min &30.74min &132.17min &33.26h &15.10x \\         
SM4&  22.52min &50.31min &35.50min &30.32min &138.65min &30.25h &13.09x \\         
SPI& 24.46min &65.25min &42.32min &35.47min &167.50min &31.26h &11.20x \\         
UART&  4.29min &16.33min &12.52min &20.55min &53.69min &20.32h &22.71x \\ \midrule        
Average &  15.61min &41.34min &24.96min &20.84min &102.75min &27.01h &\textbf{15.77x} \\    
\bottomrule    
\end{tabular}    
}    
\vspace{-5pt}     
\end{threeparttable}    
\end{table*}

Fig.~{\ref{fig:ex2}} demonstrates the verification coverage of \name\ and the other two control groups in each module. Specifically, the average code coverage of \name\ reaches 87.44\%, which is 20.96\% and 27.24\% higher than UVLLM and MEIC respectively. In terms of function coverage, \name\ achieves 89.58\%, which is 23.51\% and 28.40\% higher than UVLLM and MEIC respectively. Among them, DDR3 achieves the highest code coverage of 98.31\% and function coverage of 94.44\%. As the number of RTL code lines increases and the number of sub-modules grows, the coverage of \name\ decreases but still remains competitive. HUF, an RTL code with nearly 1,500 lines and containing ten submodules, has a code coverage of 73.07\%, however, the function coverage is 91.21\% since several code segments contribute to just one function point in this design. Moreover, the coverage of \name\ in the verification of each module is higher than that of the other two works. 
% An analysis of the results from the other two approaches reveals a mixed pattern of testbench coverage across modules, one method occasionally outperforms the other, and vice versa. This variability stems from their reliance on random stimulus generation, which inherently leads to  incomplete coverage.
This result indicates that \name\ significantly improves the test completeness in LLM-based verification. Meanwhile, \name\ also practices the theory of applying LLM to UVM-based verification, demonstrating that LLM-based testing does not have to rely solely on random stimuli.
% (\textcolor{red}{YJH: v1 complete.Need to check the description object of 'coverage', maintain consistency, and avoid module coverage, testbench coverage, and verification coverage. After determining a reasonable statement, it is necessary to fine tune the indicator description of A. When statement is correct, we need to indicate the value in fig10.})

\noindent \textbf{Result 3:}
% Table 2 presents the completion time of \name, the time consumption of each stage in the workflow, and the completion time of experienced engineers. The table shows that \name\ demonstrates competitiveness in improving the speed of functional verification. For example, in the HUF module, the speedup ratio of \name\ reaches 
% 16.67×; for code modules with lower complexity, this performance gap further increases, as \name\ shows a speedup of up to 
% 106.66× compared to human experts. By analyzing Table 2, we can find that for complex modules, the time spent on Testcase Analysis by \name\ increases, indicating that the construction of the verification testbench for complex modules often cannot be generated correctly at one time and requires multiple iterations for generation.
We recorded the time required for \name\ to complete each process, as well as the total time, comparing it with the time taken by experienced verification engineers to build  UVM testbenches. 
% As shown in Fig.~\ref{fig:flow}, Testbench Generation and Testcase Supplement are bottlenecks, so we paid special attention to the time spent in these two stages.

% As is shown in Table~\ref{Table:runtime2}, \name\ is significantly more expeditious than human verification engineers in achieving similar coverage conditions. For the HUF module, the speedup of \name\ reaches 
% 18.60×; for code modules with lower complexity, this performance gap further increases, as \name\ demonstrates a speedup of up to 45.71× compared to human experts in the AES module. The two time-consuming phases (Testbench Generation and Testcase Supplement) have average execution time of 26.90 minutes and 34.60 minutes across different modules, respectively. As shown in Fig~\ref{fig:Intro}, these two stages account for approximately 30\% and 50\% of the total workflow time. By proportional estimation, the average time spent by human experts on these two processes is 8.04 hours and 13.4 hours, respectively, representing speedups of 
% 18.49× and 23.24× for \name. This demonstrates the capability of \name\ to significantly reduce duration and its competitiveness in accelerating functional verification.
As shown in Table~\ref{Table:runtime2}, \name\ significantly outperforms human verification engineers in achieving similar coverage conditions. For the HUF module, \name\ achieves a speedup of 14.68×, while for less complex modules like DFI, the performance gap widens, with a speedup of up to 38.82×. The two most time-consuming phases, testbench generation and testcase supplement, have average execution time of 41.34 minutes and 20.84 minutes, respectively. As illustrated in Fig.~\ref{fig:Intro}, these phases account for approximately 30\% and 50\% of the total workflow time. In comparison, human experts typically spend 8.10 hours and 13.51 hours on these tasks, yielding speedups of 11.75× and 38.90× for \name\ respectively. These results demonstrate \name's potential to significantly reduce verification time and highlight its competitiveness in accelerating functional verification.

\begin{table}[t]
    \sffamily
    \small
    \belowrulesep=0pt
    \aboverulesep=0pt
    \renewcommand{\arraystretch}{1}
    \centering
     \caption{ $SR_G$ of \name -generated UVM testbenches across three iterative verification rounds. ``Gain'' indicates the percentage improvement from the previous round.}
     \label{Table:runtime3}

     \begin{threeparttable}
     \resizebox{0.88\linewidth}{!}{%
    \begin{tabular}{c| c|c  c|c c}
    %p{1.2cm} p{1.2cm}| p{1.2cm} p{1.2cm}| p{1.2cm} p{1.2cm}| p{1.2cm} p{1.2cm}}
    \toprule
    \textbf{Design} & \textbf{Round1} & \textbf{Round2} & \textbf{Gain} & \textbf{Round3} & \textbf{Gain} \\
    \midrule
    
    AES&  93.33\%& 95.56\%& 2.23\% & 95.56\% &0.00\% \\ 
    ALU&  86.67\%& 91.11\%& 4.44\% & 93.33\% &2.22\%  \\ 
    DFI&  82.22\%& 91.11\%& 8.89\% & 91.11\% &0.00\%  \\ 
    HUF&  44.44\%& 77.78\%& 33.34\% & 77.78\% &0.00\%  \\ 
    SDRAM&  42.22\%& 86.67\%& 44.45\% & 86.67\% &0.00\%  \\ 
    SHA256&  66.67\%& 84.44\%& 17.77\% & 84.44\% &0.00\%  \\ 
    SM4&  57.77\%& 82.22\%& 24.45\% & 82.22\% &0.00\%\\ 
    SPI& 64.44\%& 80.00\%& 15.56\% & 80.00\% &0.00\% \\ 
    UART&  84.44\%& 86.67\%& 2.23\% & 88.89\% &2.22\%\\ \midrule
    Average & 69.13\%& 86.17\%&17.04\% &86.67\%&0.50\%  \\
    \bottomrule
    \end{tabular}
    }
    % \vspace{-8pt}
     \end{threeparttable}    
\end{table}

\noindent \textbf{Result 4:}
% Table 3 presents statistics on the number of iteration rounds for successful generation of UVM verification testbenches for each module and the error - prone components when generation fails. The results show that only two modules can complete the verification testbench without the repair mechanism. Through our repair mechanism, the generation success rate of the verification testbench can be increased from 
% 31.11\%to 88.88\%, illustrating the necessity of the error - repair mechanism. By analyzing the error - prone components, we find that sequence\_item, driver, and scoreboard are the components with the highest error rates. Sequence\_item requires the LLM to analyze and establish constraints, driver requires the LLM to have a deep understanding of timing handshakes, and scoreboard involves the application of the TLM mechanism. This also indicates the lack of in - depth knowledge of UVM in LLM. However, through the repair mechanism of \name, which transmits error information to the LLM and provides domain - specific knowledge guidance, it can significantly enhance its ability to generate UVM components.
We evaluated the generation success rate of UVM testbenches with different iterations for the repair mechanism in \name. 
% To mitigate the impact of randomness, each module was subjected to repeated testing, ensuring a more accurate assessment of the necessity for the repair mechanism described in Testcase Analysis (Section~\ref{sc:det}). It is important to note that “multiple tests” refers to initiating \name\ multiple times; within each test, \name may involve one or more generation attempts. \zhenote{Please check.}

Table~\ref{Table:runtime3} presents the generation success rate of UVM testbenches for each module across three rounds (including one initial generation round, followed by two repair iterations) under the condition of multiple tests. The results show that with our repair mechanism, the final generation success rate is 86.67\%, reflecting an increase of 17.54\% compared to the initial generation success rate of 69.13\%. We observed that in AES and UART, the generation success rate could reach 93.33\% in the round 1, while for RTL like HUF and SDRAM, multiple rounds were needed to generate a correct UVM testbench. This reveals the model’s deficiency in understanding in-depth domain expertise regarding timing handshakes, interface protocols in IC design, and verification domains. Meanwhile, the improvement in the generation success rate between the third and second rounds is insignificant, merely 0.50\%. Further increasing the number of repair iterations would not yield better results. This validates that setting the number of iterations in the repair mechanism to 2 is reasonable, avoiding unnecessary resource waste.
% Table 3 reports the UVM testbench generation success rates for each module across three iterations, which includes an initial generation attempt and repair iterations (test conditions are defined in RQ4). 
% The results show that the application of our repair mechanism increased the overall generation success rate to 86.67\%, representing a 17.53\% improvement overall the initial success rate of 69.13\%.
% Notably, for modules like AES and UART, the success rate directly researches at the first iteration. 
% In contrast, modules like HUF and SDRAM required multiple iterations to yield a correct UVM tesetbench.
% This discrepancy highlights the model's limited understanding of domain-specific knowledge, particularly in the domain of timing handshakes, interface protocols, etc. 
% Moreover, the marginal improvement in success rate between the second and third iteration (only about 0.05\%) reports diminishing returns from addition repair iterations.
% This supports the decision to cap the number of iterations at two (in experimental setup), which optimises the resource efficiency without compromising the performance.

\begin{figure}[t]
    \centering
    \includegraphics[trim=00mm 00mm 00mm 00mm, clip, width=1\linewidth]{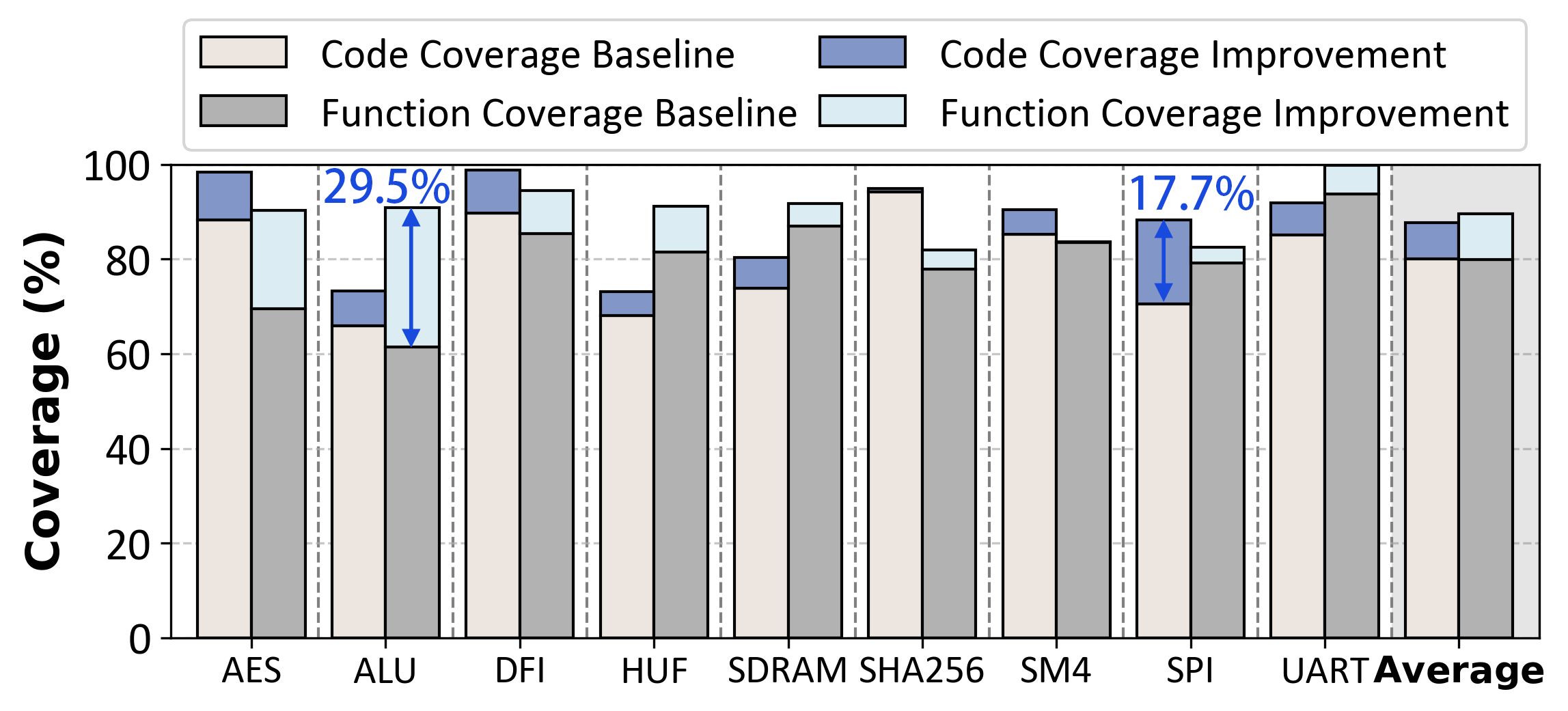} % 可根据需求调整宽度，如 0.9\textwidth
    \caption{Coverage improvement via testcase supplement}
    % \vspace{-5pt}
    \label{fig:ex5} % 便于后续引用
\end{figure}

\noindent \textbf{Result 5:}
% Fig.~\ref{fig:ex5} shows the changes in verification coverage under the action of the optimization mechanism. The results show that the optimization mechanism of \name\ can increase the code coverage and functional coverage by 7.6\% and 9.7\% respectively. Among them, for the SPI module, the optimization mechanism increases its code coverage by 37.8\%, and for the ALU module, it increases its functional coverage by 29.5\%. At the same time, we can also find that some modules show no improvement or insignificant improvement. For example, for the SM4 module, while the optimization mechanism increases its code coverage by 5.1\%, it does not improve its functional coverage. This is explainable because in actual design, there are indeed some functional points that require the joint triggering of numerous code blocks. Specifically for the current module, when the functional coverage has reached 83.50\%, the 5.4\% increase in code coverage is not sufficient to trigger more functional points. It is worth supplementing that, as shown in Table 2, the average time for Testcase Supplement is 34.6 minutes. With such a time-cost overhead, the improvement shown in Fig 11 also demonstrates the competitiveness of the optimization mechanism of \name.
We evaluated the effect of the optimization mechanism in \name\ on verification coverage by comparing the coverage with and without the testcase supplement phase. The results, presented in Fig.~\ref{fig:ex5}, highlight the improvements in both code and function coverage. Specifically, the optimization mechanism led to an increase of 7.6\% and 9.7\% in code and function coverage, separately.

Notably, the optimization is especially beneficial for certain modules. For instance, the SPI module witnesses a 17.7\% increase in code coverage, while the ALU module experiences a 29.5\% improvement in function coverage. However, some modules showed slight gains. In the SM4 module, for example, despite a 5.1\% rise in code coverage, functional coverage hasn't improved noticeably. This outcome can be attributed to the fact that, in practical designs, certain function points require the joint triggering of multiple code blocks. In the case of the SM4 module, once function coverage reached 83.5\%, an additional 5.4\% in code coverage was insufficient to trigger new function points.

It is important to note that the average time required for the testcase supplement phase was 20.84 minutes, as shown in Table \ref{Table:runtime2}. Despite the time-cost overhead, the improvements in coverage demonstrate the competitiveness and effectiveness of the optimisation mechanism.

\vspace{-3pt}
\section{Conclusion}
\label{sc:con}
\vspace{-3pt}
% In this work, we investigated LLM-based hardware desugn verification, identifying key challenges, and introduced UVMatic, a framework designed to enhance capabilities of the LLMs in this domain. By employing a hierarchical approach with systematic instructions, \name\ guides LLMs to generate UVM-compliant testbenches, achieving a 98\% pass rate and a 10-minute average execution time, marking a 110x speedup in development efficiency and a 70\% reduction in development cycle time. To address the scarcity of complex RTL benchmarks (most existing DUTs have less than 150 lines), we developed a new suite featuring modules with over 300 lines of code, better reflecting real-world verification complexity. UVMatic integrates with tools like VCS to automate the full verification workflow and incorporates a coverage optimization mechanism, enabling generated testbenches to achieve 92\% functional coverage—9x higher than state-of-the-art LLM-based verification methods. 

% Through \name, we unlock the potential of LLMs for verification, bridging the gap where manual testbench development was previously required. However, we also identify current limitations: improving LLMs’ understanding of complex hardware designs and protocols to generate precise verification code, and embedding more verification strategies and expertise into the automated generation process, will be critical future directions.
% \textcolor{purple}{data to be updated}

In this work, we explored the use of LLMs for hardware design verification and identified key challenges that hinder their effectiveness. To address these issues, we introduced \name\, a structured framework that enhances LLM capabilities through hierarchical guidance and systematic prompting. \name\ enables the generation of UVM testbench with a generation success rate of 86.67\%. With approximately 88\% coverage, the average runtime is 102.75 minutes, achieving up to 38.82x speedup in development efficiency.

Recognizing the limitations of existing benchmarks, where most DUTs are under 150 lines, we developed a new benchmark(up to 2k lines) to better reflect real-world verification demands. \name\ seamlessly integrates with industrial tools such as Synopsys VCS to automate the entire verification workflow. It also incorporates a optimisation mechanism, allowing generated testbenches to reach an average 87.44\% code coverage and 89.58\% function coverage, outperforming state-of-the-art solutions by 20.96\% and 23.51\%
respectively.

Our results demonstrate that \name\ significantly narrows the gap between LLM capabilities and the stringent requirements of industrial-grade hardware verification. However, challenges remain. Enhancing LLM understanding of hardware protocols and semantics, and embedding richer verification knowledge into the generation process, are essential directions for future work. By advancing these areas, we move closer to realizing autonomous, expert-level verification assistance powered by foundation models.

\section{Acknowledgement}
\label{sc:acknowledgement}
% We appreciate the anonymous reviewers for their helpful feedback.
%This work is supported by the National Key Research and Development Program (No. 2024YFB4405600), the National Natural
%Science Foundation of China (No. 62472086), the Basic Research Program of Jiangsu (No. BK20243042), the Key R\&D Program of Jiangsu Province (No. BE2023020-1), and the Science and Technology Major Special Program of Jiangsu (No. BG2024010).
We appreciate the reviewers for their insightful and helpful feedback.
This work is supported by the National Key Research and Development Program (Grant No. 2024YFB4405600), the National Natural
Science Foundation of China (Grant No. 62472086 and No. 92464301), the Basic Research Program of Jiangsu (Grant No. BK20243042), the Science and Technology Major Special Program of Jiangsu (Grants No. BG2024010),  and the Start-up Research Fund of Southeast University (Grant No. RF1028624005).

\clearpage
\bibliographystyle{IEEEtran}
\bibliography{Mybib}
% \begin{thebibliography}{00}
% \bibitem{b1}Xu K, Sun J, Hu Y, Fang X, Shan W, Wang X, Jiang Z. Meic: Re-thinking rtl debug automation using llms. arXiv preprint arXiv:2405.06840. 2024 May 10.
% \bibitem{b2}Hu Y, Ye J, Xu K, Sun J, Zhang S, Jiao X, Pan D, Zhou J, Wang N, Shan W, Fang X. UVLLM: An Automated Universal RTL Verification Framework using LLMs. arXiv preprint arXiv:2411.16238. 2024 Nov 25.
% \bibitem{b3} Blocklove J, Garg S, Karri R, Pearce H. Evaluating LLMs for Hardware Design and Test. In2024 IEEE LLM Aided Design Workshop (LAD) 2024 Jun 28 (pp. 1-6). IEEE.
% \bibitem{b4} Liu M, Kang M, Hamad GB, Suhaib S, Ren H. Domain-Adapted LLMs for VLSI Design and Verification: A Case Study on Formal Verification. In2024 IEEE 42nd VLSI Test Symposium (VTS) 2024 Apr 22 (pp. 1-4). IEEE.
% \bibitem{b5} I. S. Jacobs and C. P. Bean, ``Fine particles, thin films and exchange anisotropy,'' in Magnetism, vol. III, G. T. Rado and H. Suhl, Eds. New York: Academic, 1963, pp. 271--350.
% \bibitem{b6} K. Elissa, ``Title of paper if known,'' unpublished.
% \bibitem{b7} R. Nicole, ``Title of paper with only first word capitalized,'' J. Name Stand. Abbrev., in press.
% \bibitem{b88} Y. Yorozu, M. Hirano, K. Oka, and Y. Tagawa, ``Electron spectroscopy studies on magneto-optical media and plastic substrate interface,'' IEEE Transl. J. Magn. Japan, vol. 2, pp. 740--741, August 1987 [Digests 9th Annual Conf. Magnetics Japan, p. 301, 1982].
% \bibitem{b78} M. Young, The Technical Writer's Handbook. Mill Valley, CA: University Science, 1989.
% \end{thebibliography}
\vspace{12pt}

\end{document}